\documentstyle[draft,epsf]{mn}

\newcommand{\mnref}[1]{\hangindent=0.5in \hangafter=1 #1 \par}
\newenvironment{refs}{\parindent=0pt}{\parindent=1.5em}
\newcommand{\mn}{MNRAS}

\newcommand{\apj}{ApJ}
\newcommand{\pasp}{PASP}
\newcommand{\apjs}{ApJS}
\newcommand{\aaa}{A\&A}
\newcommand{\aas}{A\&AS}
\newcommand{\Msolar}{\mbox{\,$\rm M_{\odot}$}}        
\newcommand{\Lsolar}{\mbox{\,$\rm L_{\odot}$}}        
\def\gs{\mathrel{\raise0.27ex\hbox{$>$}\kern-0.70em
\lower0.71ex\hbox{{$\scriptstyle \sim$}}}}

\def\ls{\mathrel{\raise0.27ex\hbox{$<$}\kern-0.70em
\lower0.71ex\hbox{{$\scriptstyle \sim$}}}}

\title{Near Infrared Spectroscopy of the Ultracompact HII Region G45.12+0.13}
\author[S.L. Lumsden and P.J. Puxley]
{S.L. Lumsden$^{1}$
 and P.J. Puxley$^2$\\
{}$^1$ {\em Anglo-Australian Observatory, PO Box 296, Epping, NSW 2121,
Australia -- sll@aaoepp.aao.gov.au}\\
{}$^2$ {\em Royal Observatory Edinburgh, Blackford Hill, Edinburgh, EH9 3HJ,
UK -- pjp@roe.ac.uk}}

\begin{document}

\maketitle

\begin{abstract}

We present complete, low resolution $IJHK$ spectroscopy of the ultracompact HII
region, G45.12+0.13.  From the observed HI line strengths, we
 derive a near infrared extinction law that is  slightly
steeper than the average. After correction with this extinction law,
we find good agreement between the observed line
ratios of HeI, Fe$^+$, Fe$^{2+}$, S$^+$ and S$^{++}$ and the available atomic
data.   Our data show that the density within the core of G45.12+0.13 must be
at least $10^4$cm$^{-3}$.  This is consistent with the known radio structure of
the HII region and in considerable disagreement with previous work using mid
and far infrared lines.    There must also be considerable opacity in the HeI
2$^3$P--2$^3$S transition, and we show how the observed strengths of the other
HeI lines are consistent with this.   From modelling the photoionisation
structure, we find good agreement with most of the observed data if the hottest
star present has T$_{eff}\le 42000$K.  Consideration of the helium ionisation
state places a lower limit on this value so that we can also constrain
T$_{eff}\ge38000$K.  Discrepancies still exist between some of the observed and
model line ratios, but the most obvious tend to be the mid-IR
observations.  

\end{abstract}

\nokeywords

\section{Introduction}
Ultracompact (UC) HII regions are generally held to be unevolved nebula
marking the site of a young OB star, or group of such stars.  The definition
of the UC phase is usually taken to be $n_e > 10^4$cm$^{-3}$ and $r<0.1$pc.
(Habing \& Israel 1979).  Although masked in the visual by the extinction
due to the molecular cloud in which they form, they are extremely luminous
objects in both the infrared and radio wavebands.  Interest in this area
has been renewed with the completion of a sparsely sampled VLA survey of
the galactic plane by Wood \& Churchwell (1989).  They found that the numbers
of UC HII regions they detected were inconsistent with the simple picture 
in which the Str\"{o}mgren radius expands at the sound speed of the region,
and hence expands to be larger than 0.1pc after $\sim 10^4$ years
(see, eg., Osterbrock 1989).

There are several more complex models that would explain the observed numbers. 
Van Buren et al.\ (1990) have suggested that an O star may be formed with a
peculiar velocity relative to its natal molecular cloud.  The stellar wind
known to exist in main sequence OB stars  supports a bow-shock
ahead of it as it passes through the molecular cloud. The ionisation front is
trapped and the region is 
constrained from expanding by this motion.  In another scenario (e.g. Keto,
Ho \& Haschick 1987),  
material in the molecular cloud is still infalling into the HII region,
again constraining the growth of the Str\"{o}mgren radius.  Hollenbach, Johnstone
\& Shu (1993) have proposed that a remnant circumstellar accretion disk may
replenish the nebula with new material, hence intercepting much of the far UV
light from the exciting star.  
Dyson, Williams \& Redman (1996) advanced a somewhat similar model, in which 
the interaction of the stellar wind with a clumpy
molecular cloud can lead to mass loading in the wind, creating the same
effect as in the Hollenbach et al.\ model.
All of these models can satisfy the lifetime
arguments, but all differ in the other physical properties of these regions
(eg.\ the dynamical structure).

The advent of near-infrared array spectrometers has made it possible to obtain
spectra of bright nebular sources with a signal-to-noise that would have been
impossible before.  Although compact HII regions have been observed widely in
the near infrared in the past (see below), only very bright features with high
equivalent width have been detected.  There are also many intrinsically fainter
features that could be present.  By studying these we can accurately determine
the extinction to the HII region, the conditions present in the regions
of different ionisation, examine the impact of the large opacity in the
HeI triplet 1.083$\mu$m $2^{3}$P--$2^{3}$S transition on the weaker HeI
lines present in the near-IR, and perhaps find transitions that may
indicate the presence of bow-shocks (coronal lines are expected to be
detectable in these models) or remnant accretion disks (strong neutral
transitions or perhaps CO emission).

We have therefore carried out a series of observations of one of the most
luminous sources in the VLA survey of Wood \& Churchwell, G45.12+0.13. This
ultracompact HII region has been well studied at many wavelengths, both prior
to and since the Wood \& Churchwell survey.  There have been observations of
the hot excited molecular gas around the region (eg.\ Churchwell, Walmsley \&
Wood 1990, Cesaroni et al.\ 1991), of the continuum emission both at mm and
sub-mm wavelengths (Hoare, Roche \& Glencross 1991, Wood, Churchwell \& Salter
1988, Wood et al.\ 1988), and of the radio recombination lines (Garay, Reid \&
Moran 1985, Churchwell et al.\ 1990).  Wood \& Churchwell classify this source
as `cometary' (with a bright bow-shaped ionisation front in the radio map),
but, as Colgan et al.\ (1991) note, it can also be modelled as a core-halo
source, since there is considerable extended emission beyond the bright core
observed by Wood
\& Churchwell (Matthews et al.\ 1977, Wink, Altenhoff \& Mezger 1982).
Herter et al.\ (1981) collected the results of all the near-infrared 
spectroscopy carried out on this source with the first generation of low
resolution instruments.  Since then it has largely been overlooked in this
waveband, though it has been observed in the mid and far-infrared using both
IRAS and the KAO (Simpson \& Rubin 1990, Colgan et al.\ 1991).  

The best models of G45.12+0.13 show a core 
with a density $n_e>10^4$cm$^{-3}$, and a lower density halo that
may reflect the fact that the ultracompact HII region is embedded in a larger
ionised zone (eg.\ Colgan et al.\ 1991).  
Wink et al.\ (1982) determine the electron temperature to be
T$_e=8000\pm2000$K, in agreement with the data from Wood \& Churchwell.

This paper presents our results for the 0.85--2.5$\mu$m region.  In section 2,
we detail the observations made, and the data reduction techniques involved.
In section~3, we present the spectra of G45.12+0.13, and identify spectral
features.  In section~4, we discuss the properties of the identified lines, and
from these discuss the possible conditions within the nebula.  Finally, in
section~5, we present the conclusions of this work.

\section{Observations}
The data for this paper were accumulated over a period of time, under varying
conditions.  The initial observations were made at UKIRT with the common user
infrared array spectrometer CGS4 (Mountain et al.\ 1990).  CGS4 used a SBRC
InSb 62$\times$58 array sensitive from 1--5.5$\mu$m. In all cases the spatial
resolution was 3$''$ per pixel, and the slit-width was also 3$''$.  The slit
was oriented north-south, across the peak of the infrared emission. We summed
over five rows to obtain total fluxes for the object, so that,  for the data
presented here, each spectral point corresponds to a 3$''\times15''$ pixel
spatially.   Since the slit in CGS4 is only one pixel wide, the array is
stepped to provide properly sampled spectra.  Our data are sampled 4 times over
2 pixels. Details of the dates of observation, the wavelength range covered,
the resolution, the actual on-source integration times and the airmass at the
time of observation are given in Table 1.

We also obtained a slightly lower signal-to-noise $IJHK$ spectrum with the
common user infrared imaging spectrometer IRIS on the AAT.
IRIS uses a Rockwell HgCdTe 128$\times$128 array sensitive out to
2.5$\mu$m.  A detailed description of IRIS can be found in Allen et al.\
(1993).  In this case the slit width was 1.5$''$ and we summed over four pixels
(3.2$''$) spatially.  Details of these observations are given in Table 2.

In addition, we will also refer to data on this object taken for other projects.
These include a Br$\gamma$ image of G45.12+0.13 taken with IRIS, with a spatial
resolution of 0.25$''$/pixel, and a high resolution spectrum taken with CGS4
and the echelle grating (resolution$\sim16000$) centred on Br$\gamma$ (see
also Lumsden \& Hoare 1995).

For data obtained in 1991, conditions were non-photometric and the seeing was
poor. The atmospheric standard used was BS7332, an A2V star: it was observed
both before and after the observations.  To compensate for the airmass
corrections, we averaged the two observations and used that as the standard for
the object. Further observations of two K0III stars showed that this resulted
in a reliable atmospheric standard.   For the 1992 observation, the atmospheric
standard used was SAO104728, an A2 star. The UKIRT flux standard, GL748 was
also observed, and provides a reliable check of the atmospheric cancellation
short of 2.25$\mu$m (there are several absorption features in the flux standard
beyond this).  For the 1993 data, BS7332 was used as the atmospheric standard,
and GL748 was again observed to provide accurate photometry and to enable
spurious features in the final spectra caused by intrinsic absorption features
in the standards to be allowed for. Conditions were photometric for both the
1992 and 1993 observations, though the seeing was slightly poorer for the
latter.  The last segment of data taken on 1993 April 4 (from 2.3--2.5$\mu$m)
 is considerably poorer
as it was taken during sunrise when the night sky lines were increasing
rapidly in flux.
Wavelength calibration was achieved using either krypton or argon arcs
for the 1--1.2$\mu$m, 1.6--2.1$\mu$m and 2.3--2.5$\mu$m data, and bright sky
lines for the 1.2--1.6$\mu$m and 1.9--2.3$\mu$m data (line wavelengths were
taken from the list of Oliva \& Origlia, 1992).  For the IRIS data the G dwarf
BS7644 was used as as both a flux and atmospheric 
 standard, and wavelength calibration was by means of an
argon arc lamp.

All UKIRT data were taken by nodding along the slit to provide a sky frame
taken at the same position on the chip as the object.  The object was moved
between 27 and 30$''$ on the chip in the process.  This procedure aids
considerably in removing residual sky features, which cancel when the positive
and negative beams are added at the end of the data reduction.  For the IRIS
data, the same procedure was used but the nod distance was considerably less
since the slit is only 12$''$ long.  Given the highly peaked spatial profile of
G45.12+0.13, this method gives reliable core fluxes.

The reduction procedure for the CGS4 data followed a standard procedure. First,
the sky frames were subtracted from the corresponding object frame.  A
polynomial fit in the spectral direction  was then made to remove the residual
sky lines. The rows containing the positive and negative beams were then
removed.  Since CGS4 fully samples the resolution element by stepping the
array, variations in seeing or poor photometric conditions can result in the
four sub-spectra having different recorded count levels.  This is corrected for
by folding the combined spectrum over a period of 2 pixels to provide a
`ripple-spectrum',  which the object spectra are then divided by.  The
resultant data were then coadded, leaving for any object two summed spectra
corresponding to the original positive and negative beams.  We interpolated
over hydrogen absorption lines in the A-type standards (the only feature
present in their spectra) by fitting Lorentzian line profiles.  The separate
halves of the data were then divided by the same beam from the standard (which
corrects for differences observed in the overall raw count level in the two
beams), and then coadded.  Finally spectra were then corrected for the
black-body shape of the standard, and flux calibrated where appropriate.  Data
from non-photometric nights were calibrated by forcing the continuum in the
overlap between spectra to match.  In doing this, we have assumed that our best
photometry occurs in the 1.6--1.8$\mu$m segment of the final spectra.  Small
differences ($<5$\%) found in the continuum level from flux calibrated regions
were also removed in the same way.  Unfortunately, this procedure is
inapplicable in the overlap between $J$ and $H$, where the atmospheric
transmission is very poor.   For this reason we obtained the IRIS spectra as
well, and the overall shape  agrees well in both data sets.   We are confident
that each segment is calibrated accurately to within 10\% of its neighbour.

IRIS echelle data is considerably different from CGS4 data in that the echelles
are cross-dispersed to obtain complete coverage of $IJ$ or $HK$ in a single
exposure.  Therefore corrections for curvature of the echelle orders have to be
applied before raw spectra can be extracted.  Furthermore, since the slit in
IRIS is fully sampled by the chip there is no need to correct for variable
transmission between sub-spectra as was the case with CGS4.
 Lastly, given the short length of the slit it was impractical to
model residual sky emission, and we were forced to rely on nodding along the
slit to remove the sky features.  In other respects the data were treated in
the same fashion as the CGS4 data.

Allowance must also be made for the deep atmospheric absorption features
between 1.85 and 2.08$\mu$m and between 1.35 and 1.45$\mu$m, which
are primarily caused by atmospheric water vapour and CO$_2$.  The difference
between a good high altitude site such as Mauna Kea and a lower
altitude site such as Siding Spring is clearly evident in the much
poorer spectra obtained through these regions with IRIS.  In particular, the
CGS4 data acquired on the night of 1993 April 4 are almost photometric
through the gap between H and K windows, showing that the column density of
water vapour was especially low.  However, despite this, we assume that
the lines that fall in these wavelength ranges are of essentially lower
photometric accuracy.  The quoted errors from the line fits described below are
therefore likely to be very much a lower bound to the true error.

For one line we have attempted to model the atmospheric response to remove this
uncertainty.  The HeI 2$^1$S--2$^1$P transition at 2.058$\mu$m falls in the
midst of CO$_2$ absorption.  The intrinsic width of the absorption features are
very narrow, as are the intrinsic line widths of the HII region. Therefore, at
different times according to the object's velocity with respect to the earth,
more or less of this flux can be absorbed by the CO$_2$ features. We therefore
applied the corrections described in Doherty et al.\ (1993) to derive a better
estimate of this line.  We observed the HeI line on two separate occasions with
CGS4.  The respective corrections we derive are 1.152 and 1.252 for the 1992
November 15 and 1993 April 4 CGS4 data respectively.  After correction, the
difference in the observed line flux between the two nights is less than their
respective errors.  We therefore find this procedure to be reliable and quote
the mean of these fluxes in Table 3.

\section{The Structure of G45.12+0.13}
In Figure 1, we display a grey scale plot of the Br$\gamma$ image obtained with
IRIS.  The pixel scale is 0.25arcsec/pixel, and the seeing was $\sim$1arcsec
at K.  The image is not flux calibrated.
The underlying continuum has not been subtracted but, as will be shown
later by the spectra, this continuum is largely free--free dominated in any
case.  We can compare this image with the 15GHz radio map presented by Wood \&
Churchwell (1989).  The general outline shown by the radio data maps
well onto the higher contours of the IR data.  
However, the Br$\gamma$ map shows considerably more extended structures around
these.  

The similarity of the two maps indicates that the extinction across the HII
region is largely uniform.  This is confirmed by the lack of evidence for
variations in the extinction from our long slit spectroscopy (see section 5.1).
However, for a proper comparison we must also ensure that the radio continuum
emission is not optically thick.  As shown by Wood \& Churchwell, many UC HII
regions are still marginally optically thick even at 15GHz, and although there
is no evidence for this in G45.12+0.13 when averaged across the whole nebula,
it may still be true near the core.  To assess the likelihood of this we
compared the cross-sections through the peak of the emission in both the radio
and the near-IR.  This reveals very similar profiles in both: the radio core
can be fit reasonably by a Gaussian of full width at half maximum of 1.1$''$
and the IR core by a Gaussian with 1.75$''$.  Taking account of the seeing at K
and possible positioning errors in taking these sections, these profiles are
consistent.  This implies that any residual opacity in the radio must be small,
as expected.  We can also show this by comparing the the peak flux per beam
quoted by Wood \& Churchwell (1989) at 5 and 15GHz. From this it seems very
likely that $\tau_{\rm 15GHz}\ll 0.1$.  The slight differences in beam sizes
and the slight misalignment of the peak locations at the two frequencies do not
allow us to say more than this however.  It is clear however that the opacity
at 15GHz is sufficiently small that we can ignore it in what follows, and we
can treat the 15Ghz radio map and the near-IR spectra and image 
as tracing the same structures.

Lastly, we note that the total structure of G45.12+0.13  is not remarkably
cometary, as Wood \& Churchwell have claimed, but is rather more amorphous. 
This is consistent with the model fit of Colgan et al.\ (1991), and it seems
unlikely that the bow-shock hypothesis of Van Buren et al.\ applies to this
source on this basis alone.

\section{The Spectrum of G45.12+0.13}
The CGS4 spectra are shown in Figure 2, and lines identified are given in Table
3.  The region from 0.9--1.0$\mu$m is taken from our IRIS spectra, scaled by
the relative strength of Pa$\beta$ in the CGS4 and IRIS data to give an
indication of the strength of the [SIII] lines.  
The weaker lines are more readily seen in Figure 4 where the strong lines
have been suppressed to show the continuum in more detail.
 We do not otherwise show the
IRIS spectra separately, but the line strengths measured are also given in
Table 3.    

The lines were fitted with either a single triangle, or in the case of blends,
two triangles. This is the correct function to fit for an unresolved line with
CGS4 (or IRIS) and was found to be satisfactory for virtually every feature.
Where complex blends of more than 2 lines occurred, trying to fit more than two
triangles to the observed profile was largely unsuccessful (though we did adopt
this approach for one blend -- see section 5.4). We give wavelengths as
measured from these fits, but, with the exception of highly blended lines, the
errors on these fits are not a realistic measure of the true observational
error: for the signal-to-noise achieved in these observations, the typical
`true' error on the observed wavelength is approximately a tenth of a
resolution element.  Therefore, we quote this error, $\delta\lambda$, unless
the actual error on the fit is worse than this (which only occurs with highly
blended lines).  Note that the wavelength calibration beyond 2.3$\mu$m has an
offset present.  We have checked the line identifications against the known
wavelengths of the HI Pfund series to correct for this. Marginally detected
lines are included in the list: caution should be used in interpreting such
features. Lines which are sufficiently blended that they cannot be resolved are
marked by $bl$ after the wavelength as well.  Where we cannot deblend these
features with any confidence, the flux is also marked with a $bl$ and indicates
that this is the total flux in the blend.

The identification of each feature is also given in Table 3.  Lines for which
we are unsure of the identification are marked by a ? in this column.   As one
might expect, most of the bright features arise from hydrogen recombination
lines.  The only lines of comparable strength are the HeI 2${}^3P$--2${}^3S$
1.083$\mu$m metastable triplet `ground-state' and the singlet
2${}^1P$--2${}^1S$ 2.058$\mu$m transition and the [SIII] lines at 9535\AA\ and
9072\AA.   Amongst the weaker features, the most common are the other low-$n$
HeI recombination lines not coincident with hydrogen lines, and many
transitions from the low lying [FeII] multiplets.

Line fluxes are taken from the fitting procedure, as are the quoted errors. It
should be noted that our data spans regions of very low atmospheric
transmission (primarily the gap between the $J$ and $H$ windows, between $H$
and $K$ and the region from 1.1--1.2$\mu$m).  Our IRIS data are unreliable in
these regions and no line strengths are quoted.   The fluxes quoted for lines
in our CGS4 data should also be treated with caution in these regions, and true
errors on the line fluxes are unlikely to be well represented by the fitting
process.   The fluxes listed are all scaled relative to Pa$\beta$, and all have
been corrected for extinction using the extinction curves derived in 
section 5.1.  
The original observed line fluxes
can be regained by using the appropriate parameters in the extinction curve,
the value of the Pa$\beta$ flux after correction for extinction given in 
Table 3 and the strengths  of the other lines given there also.
For highly blended
lines the quoted fluxes may also be in error.  The total flux in the blend
however should be reliable.  The `true' error in any given line can
be estimated by comparing the differences in the ratios of the 
HI series between the IRIS and CGS4 data, though it should be noted that this
may reflect real changes across the face of the nebula since the slit 
aperture and
position angle is different for the CGS4 and IRIS data.

\section{Physical Conditions within G45.12+0.13}
\subsection{Extinction}
Typically the near infrared extinction law is derived by assuming that the dust
opacity follows a power law: ie.\ if $\kappa_{\rm dust} \propto
\lambda^{\alpha}$, then the differential optical depth between two lines,
\[ \Delta\tau_{12} = -\ln 
 \frac{\left[I(\lambda_{2})/I(\lambda_{1})\right]_{\rm obs}} 
           {\left[I(\lambda_{2})/I(\lambda_{1})\right]_{\rm theory}} ,\]
is equivalent to 
\[ \Delta\tau_{12} 
= \tau_{\lambda_{1}}\left[(\lambda_{2}/\lambda_{1})^{\alpha} - 1
\right].\]
Obviously, in the case where we know $\alpha$ and have a measurement of a
line (or free-free processes) where $\tau=0$, we can derive $\tau$ at the
wavelength of the line absolutely.

The extinction can be derived in several different ways.  First, we can compare
the observed line fluxes with the radio continuum, which arises from free-free
emission.  
The expected radio emission can be calculated as shown by Rubin (1968),
and the expected line flux can also be derived using the calculations
of Hummer \& Storey (1987).  Combining these we get in the general case
(assuming the nebula is ionisation bounded and there is no He$^{++}$,
which are reasonable in this case)
\[
F_{Br\gamma, theory} = 2.45\times10^{13} S_\nu \nu^{0.1} T_4^{0.35}
j_{Br\gamma} {\rm Wm}^{-2},
\]
where $S_\nu$ is given in Jy, $\nu$ in GHz, $T_4$ is in units of $10^4$K
and $j_{Br\gamma}$ is in Wm$^{-3}$ and is taken directly from Hummer
\& Storey.  For $n_e=10^5$cm$^{-3}$ and $T_e=7500$K,
$j_{Br\gamma}=4.64\times10^{-28}$Wm$^{-3}$.  

We estimated the total radio flux within the
infrared slit from the Wood \& Churchwell radio maps.   However, since we
observe hydrogen emission beyond the lowest contour in their radio maps, it is
clear that their data is not sensitive to low level large scale emission (the
VLA configuration used is insensitive to structures on scales larger than
10$''$).  Therefore, to minimise the errors in carrying out this procedure, we
have only used data corresponding to the emission peak: in practice this is
defined by the 0.8$''\times1.5''$ aperture of IRIS (1$\times$2 pixels) for
which we observe a Br$\gamma$ flux of (1.6$\pm$0.02)$\times10^{-15}$Wm$^{-2}$. 
We estimate the total radio flux within this aperture as 525mJy at 2cm
(assuming a Gaussian profile for the core of the radio emission with
$\sigma\sim0.5$arcsec and the peak flux given by Wood \& Churchwell 1989). 
Hence combining this
with the 15GHz data of Wood \& Churchwell gives:
\[ F_{Br\gamma, theory} = 7.1\times10^{-15}{\rm Wm}^{-2}.\]
Allowing for errors in the radio flux (resulting
both from our assumptions and from possible mismatch between the centreing of
the apertures) and the electron temperature, the total error in this flux is
$\sim$30\%.  Hence $\tau_{{\rm Br}\gamma}=1.49$, with a
likely error of $\sim0.3$. We have assumed in this that the 15GHz emission is
optically thin: free-free opacity will result in a larger value of $\tau_{{\rm
Br}\gamma}$ than given here.  However, for the estimate of the free-free
opacity at 15GHz given in section 3, the difference is essentially within the
error quoted.

Secondly, we can compare the hydrogen line ratios themselves.  This is the
method recommended by Landini et al.\ (1984) for measuring the extinction law
in a more general case.  It has the advantage that we can derive both $\tau$
and $\alpha$ since we have many recombination line measurements. The model data
has been taken from the case B recombination line predictions of Hummer \&
Storey (1987), where again
we have used $T_e=7500$K and $n_e=10^5$cm$^{-3}$ (the
values are only slightly sensitive to density within the range in question but
are sensitive to electron temperature).  The observed fluxes for the
Pfund, Brackett and Paschen series lines observed with CGS4 are plotted in
Figure 3a, together with the model values derived from Hummer \& Storey.
Figure 3b  shows similar data derived from our IRIS observations.

With these assumptions we find $\tau_{{\rm Br}\gamma}=1.45\pm0.06$ and
$\alpha=-2.01\pm0.06$ from our IRIS data, and $\tau_{{\rm
Br}\gamma}=1.35\pm0.04$ and $\alpha=-2.01\pm0.05$ from our CGS4 data.  In
carrying out the fits we have excluded 
those points 
lying in regions of poor atmospheric transmission, and those which are
blended with other features.  The fits have reduced $\chi$-squared values of
0.9 and 2.2 respectively.  The better fits obtained with IRIS highlight the
problems in accurately matching segments of CGS4 data taken at different
epochs.  In particular, the data between 1.5$\mu$m and 1.6$\mu$m can be seen to
be slightly offset from the longer wavelength CGS4 data.  These fits give a
predicted visual extinction of A$_V=23-25$.  The exact agreement between
$\tau_{{\rm Br}\gamma}$ from our IRIS measurements and from the radio data must
be seen as being somewhat fortuitous given the assumptions we have made. All
values of $\alpha$ and $\tau$ that we derive are also compatible with our
non-detection of the higher members of the Paschen series.  Since the values of
$\tau_{{\rm Br}\gamma}$ agree within the errors for the two data-sets we
conclude that there is no evidence for strong deviations in the extinction
across the face of the nebula.  Similarly, we see no evidence that the value of
$\tau_{{\rm Br}\gamma}$ changes along the slit for our CGS4 data.  

Since the interpretation of the HeI line spectrum is complicated (see section
5.2), it is inadvisable to use HeI line ratios in deriving estimates of the
extinction.  The one exception to this is
the 4$^3$P--3$^3$D/4$^3$P--3$^3$S pair which have a common upper level.  
 However, since the former line falls in a region of poor
atmospheric transmission, the result again has a large error.  Using the IRIS
data, we derive $\tau_{{\rm Br}\gamma}=1.55\pm0.2$, assuming $\alpha=-2.01$.
This error is likely to be somewhat conservative, given the comments above on
the photometric accuracy of any line that falls between 1.8 and 2.06$\mu$m.
However, the result is consistent with the assumed slope.  Using the CGS4 data,
the derived $\tau_{{\rm Br}\gamma}=1.38$, again consistent with our earlier
result.

Lastly, we can compare line ratios of other ions.  The ratio of [FeII]
1.257$\mu$m to [FeII] 1.644$\mu$m is a potentially powerful indicator of
extinction since the two lines have a common upper level.  In practice, since
the 1.644$\mu$m line is heavily blended with Br12, the uncertainty introduced
is large.  We discuss the ratios of these two lines further in section
5.3.  One other potential indicator is the ratio of the [SIII]
${}^1$D$_2-{}^3$P$_1$ and ${}^1$D$_2-{}^3$P$_2$ lines at 9072\AA\ and 9535\AA,
since these also share a common upper level.  As shown in section 5.4, the
derived extinction from these lines agrees well with that derived from the HI
lines.

Although the values of $\tau_{{\rm Br}\gamma}$ measured by the two instruments
are statistically equivalent, it is possible that there are variations in the
measured $\tau$ due to variations in the optical depth through the dust along
the line of sight.  Natta \& Panagia (1984) outline the expected observed
versus intrinsic extinctions in the general case of an inhomogeneous screen.
For the small values of $\tau$ observed, a change of $\sim$10\% in these values
as measured between a small aperture (essentially closer to the `intrinsic'
value) and a large aperture (covering most of the source as is the case for our
CGS4 data) is entirely possible if we imagine the dust screen as existing in a
thick shell around a mostly spherical HII region (equivalent to the $n=1$ case
of Natta \& Panagia).  However, plotting  $\log\tau$ (derived by comparing the
IR fluxes with those expected from the radio for all $\lambda$) as a function
of $\log \lambda$ as they do (cf their figure 10), shows no evidence for a
turnover towards a flatter slope at shorter wavelengths, which would be the
characteristic for such behaviour. Since we do not see any of the optical HI
lines however, it would be difficult to fully test the `clumping' model from
the data available to us at present.  It is clear though that the slope we
measure should only relate to the intrinsic extinction present.

Previous extinction estimates for this source have ranged from
A$_V=10\rightarrow30$, as discussed by Hoare et al.\ 1991. They express concern
that the hydrogen recombination line strengths may be affected by stellar winds
based on the broad HI lines detected in the high resolution observations of
Tanaka et al.\ (1985).  However, a velocity resolved spectrum of the Br$\gamma$
line shows that the intrinsic width is $\sim45$kms$^{-1}$, and we therefore
reject the Tanaka et al.\ result as incorrect.  We conclude that G45.12+0.13
has an extinction curve that is somewhat steeper than that found in optically
visible nebulae such as reported by Landini et al.  In what follows, we assume
the use of the extinction curve derived from the HI series to deredden our
data, with $\alpha=-2.01$ for both the IRIS and CGS4 data, and with $\tau_{{\rm
Br}\gamma}=1.45$ for the IRIS data and $\tau_{{\rm Br}\gamma}=1.35$ for the
CGS4 data.

\subsection{Helium Recombination Lines}
Basic recombination line strengths for the infrared HeI lines are presented by
Smits (1991a,b), for the case of 10\% helium abundance.  There are several
effects that are not included in this model however.  These include collisional
excitation from the 2$^3$S metastable level to levels higher than $n=3$,
photoionisation of the 2$^3$S level (Clegg \& Harrington 1989), the inclusion
of the role of the optically thick $n^3$P-2$^3$S series (Robbins 1968) and the
role of dust and hydrogen in absorbing scattered 2$^1$P-1$^1$S photons
resulting in a breakdown of the case B approximation to a small extent (Doyon,
Puxley \& Joseph 1992, Shields 1993).  Of these, we can discount
photoionisation as an important factor from the work of Clegg \& Harrington.
Their model of the low excitation compact planetary nebulae IC418 is a
reasonably close approximation to the situation in G45.12+0.13 and, as they
show, collisional excitation of the lower singlet levels dominates over
photoionisation as a means of depopulating 2$^3$S. For the other possible
contributing effects, since we have accurate fluxes for a number of the low
lying HeI transitions, we can directly test for their presence.

Table 4 gives the ratios of the observed HeI lines with the
4$^3$D-3$^3$P line.  This latter 
line arises from the same state as the more
commonly used 4471\AA\ 4$^3$D-2$^3$P transition that is well known to be
largely unaffected by either collisions or significant line opacity (eg.\ Clegg
1987).  
 Table 4 also gives the theoretical ratios as taken directly from
Smits (1991b): with the exception of the 2$^{1,3}$P--2$^{1,3}$S transitions,
these represent the values derived from direct recombination and cascade
processes only.  The values are for T$_e=10^4$K and 5000K, and
$n_e=10^4$cm$^{-3}$.  Changes in the electron density have a negligibly small
effect on the values tabulated by Smits for all line ratios except
those involving the 2P--2S transitions.  Smits includes collisional excitation
processes for these lines in his predictions.  The electron temperature for
G45.12+0.13 will lie somewhere between these limits as noted previously.

The measured flux for the 4$^1$P--3$^1$S line assumes that the contribution
from HI Br22 can be derived from the measured Br21 flux and the ratio of
Br22/Br21 as derived from Hummer \& Storey for T$_e=7500$K.
The 5$^{1,3}$F--3$^{1,3}$D
lines are also heavily blended with both Pa$\beta$ and the 
a$^4$D$_{3/2}$--a$^6$D$_{3/2}$ transition of
[FeII]: there is a considerable difference seen between the ratio observed with
CGS4 and that with IRIS, so we conclude that the quoted flux of this line is
unreliable.  The 7$^3$P--4$^3$S line at 1.745$\mu$m is coincident with the 
a$^4$D$_{1/2}$--a$^4$F$_{3/2}$
[FeII] transition: again the ratio is likely to have a larger error than that
quoted.  Lastly, the 5$^3$P--3$^3$D line is blended with both the 
  [FeII] a$^4$D$_{3/2}$--a$^6$D$_{1/2}$ and a$^4$D$_{5/2}$--a$^6$D$_{5/2}$
transitions and may also have a larger error than quoted.  The
2$^{1,3}$P--2$^{1,3}$S lines are discussed in greater detail below.  The
4$^1$P--3$^1$S agrees well with the theory, after making the correction for
the HI contribution to the blend, but 
the 4$^1$D--3$^1$P line at 1.909$\mu$m does not.  It falls on the edge of
the deep atmospheric OH absorption band however.  
The 5$^3$D--3$^3$P transition also
agrees well with the theory.  However, the 4$^{1,3}$S--3$^{1,3}$P blend shows
some enhancement compared to expectation, and the 4$^3$P--3$^3$S,
5$^3$P--3$^3$D, 7$^3$P--4$^3$S and 4$^3$P--3$^3$D are all significantly
enhanced over the expected recombination values.

The only mechanisms other than recombination for populating the higher $n$
levels are opacity in the $n^3$P-2$^3$S series and collisions from 2$^3$S.
Since the observed ratios and theoretical ratios are most discrepant for those
lines which originate from $n^3$P levels this is strongly suggestive of high
opacity.  Collisions would preferentially enhance $\Delta L=0$ transitions more
than any others.  We can quantify the relative importance of collisions to
these higher $n$ levels in the same manner as Clegg (1987).   For the case of
collisional excitation of 4$^3$S from 2$^3$S, we find the collisional
excitation rate to be $\sim10^{-12}$cm$^{3}$s$^{-1}$.  The effect this
has on the 4$^{3}$S--3$^{3}$P transition can then be calculated by considering
the comparative importance of recombinations and collisions to populating
4$^{3}$S. The ratio of excitations by collisions to those by recombinations is
simply just $N(2^3S) k_{eff} / N(He^+) \alpha_{eff}$.  The relative population
of the 2$^3$S level is taken from Clegg (1987), $\alpha_{eff}$ is derived from
the values in Smits and $k_{eff}$ is derived from the collision strengths of 
Sawey \& Berrington (1993), scaled by the branching ratio of the
4$^{3}$S--3$^{3}$P and 4$^{3}$S--3$^{2}$P transitions.  This implies that only
$\sim3$ in every thousand 2.1128$\mu$m 4$^{3}$S--3$^{3}$P photons arise from
collisions.  

The optical depth in the core of the 2$^3$P--2$^3$S 1.083$\mu$m line can be
estimated crudely by assuming a typical density of 10$^4$cm$^{-3}$, a core
radius of 0.02pc (determined from our fit to the radio data in section 3
above), and purely thermal line broadening: from this we derive
$\tau(1.083)\sim400$, or $\tau(3889)\sim20$.  Such values would imply that even
the $7^{3}$P--2$^{3}$S line has an opacity $\sim1$.  Robbins (1968) gives the
expected line enhancements for a few of the stronger infrared lines as well as
the dominant optical transitions.  We have assumed the case of purely thermal
line broadening, even though we know this to be incorrect from the observed
line widths in G45.12+0.13, since we do not know what component of the observed
line broadening is due to real velocity shifts (which would have a major impact
as noted by Robbins) and what is due to turbulence (which would have less of an
effect).  

From these data we can derive expected line enhancements for all the lines that
do not originate from $n^3$P. The ratio of the enhancement over the original
recombination line flux is the same for all transitions from any given state. 
Therefore, to calculate the enhancement in the $4^3$S--$3^3$P transition, we
only require to know the enhancement in the $4^3$S--$2^3$P transition which is
given by Robbins. In this way we also correctly account for cascades from
higher levels into $4^3$S.  This gives the enhancement as $\sim33$\%, in
agreement with what is observed. Similarly, those lines arising from 
$n^3$D should show negligible enhancement ($<$5\% for both 4$^3$D-3$^3$P 
and  5$^3$D-3$^3$P), again in agreement with what is observed.

For the lines that do arise from $n^3$P, we adopted an alternative approach,
since the quoted line fluxes in Robbins relate to the optically thick lines. If
we assume that all the light in the $n^3$P-$2^3$S lines is redistributed into
other lines arising from the same $n^3$P state (as a first approximation), then
the ratio of the extra flux to the original recombination line flux is simply
given by the ratio of all the branching ratios from that state to the branching
ratios of all those transitions except the $n^3$P--2$^{3}$S one.  We can
compare the numbers we derive with the two cases given by Robbins for near-IR
lines. This is a reasonable approximation in this case since cascades from
higher optically thick transitions are less important in feeding  these
transitions than for those lines arising from outside $n^3$P. For $4^3$P this
leads to a predicted enhancement of a factor of 4.6, for $5^3$P a factor of 3.4
and for $7^3$P a factor of 2.6. By comparison, Robbins quotes an enhancement
for 4$^3$P--3$^3$S of 4 and for 4$^3$P--3$^3$D of 3.5 for the case where
$\tau(3889)=75$. Although our crude estimates  are somewhat too high, they are
clearly close to the `true' estimates when the line is truly optically thick.
The actual observed enhancements for the transitions originating from $4^3$P
and $5^3$P are $\sim4$ in reasonable agreement with what is observed. The
observed enhancement for the $7^3$P--$4^3$S line is $\sim7$, much larger than
expected.  This is the only truly discrepant line however, and as noted above,
is completely blended with an [FeII] transition.

As shown by Clegg (1987), at high electron density, the  $2^3$P--$2^3$S line is
dominated by collisional excitation (by a factor of $\sim5$ from his data). 
Collisions between the $n=2$ levels are explicitly allowed for  by Smits, and
we also find substantial agreement with his data for the adopted value of $T_e$
and $n_e$.   The $2^1$P--$2^1$S line is also considerably modified from its
pure recombination value.  In this instance resonance scattering of the 584\AA\
$2^1$P--$1^1$S line can lead to additional $2^1$P--$2^1$S photons.   However,
$2^1$P--$1^1$S photons can ionise hydrogen and be destroyed by dust as well.  
In addition, the population of the 2$^1$S level can also be raised by
collisional excitation from $2^3$S.   As shown by Shields (1993), this problem
must be treated using a full radiative transfer model to correctly model the 
expected $2^1$P--$2^1$S line flux as a function of the stellar effective
temperature.  The major difference between that work and more  simplistic
calculations (eg.\ Doyon et al.\ 1992) is that Shields predicts the ratio of
the $2^1$P--$2^1$S with Br$\gamma$ decreases above 40000K rather than
saturating when the H$^+$ and He$^+$ zones fully overlap.   We have already
demonstrated an example of such behaviour in a previous paper (Lumsden, Puxley
\& Doherty 1994). Our observations here are in good agreement with the expected
maximum value for the $2^1$P--$2^1$S/Br$\gamma$ ratio (we assume the actual
maximum value is somewhat better defined by Doherty et al.\ 1994 than by
Shields, since the only case he considers with high density appears to be off
the scale on his Figure 1(d)).  This leads us to conclude that the exciting
stars in G45.12+0.13 must be cooler than $\sim$42000K.  Lastly, the strength of
the $2^1$P--$2^1$S line indicates that the 2$^3$S level is predominantly
depopulated by collisions, and not photoionisation, in agreement with the model
results of Clegg \& Harrington (1989).

As a further test of the ratio of He$^+$/H$^+$, we can use the 4$^3$D--3$^3$P
transition, since it arises solely from recombination and is straightforward to
interpret since the 4$^3$D--3$^3$P/Br$\gamma$ ratio does saturate once the
H$^+$ and He$^+$ zones overlap (cf.\ the discussion of the optical HeI lines in
this context in Doherty et al.\ 1995). The ratio of this line with Br$\gamma$
gives $0.12\pm0.01$ from the average of our CGS4 and IRIS data. The theoretical
value for T$_e=7500$K and $n_e=10^4$cm$^{-3}$ is
1.32$n_{He^+}/n_p$V$_{He^+}/$V$_{H^+}$.  Increasing the density by an order of
magnitude causes a drop of $\sim$10\% in the theoretical ratio, as does raising
T$_e$ to 10000K.  Assuming a nominal helium abundance by number of 10\% of
hydrogen implies that our observed ratio can only be satisfied if the hydrogen
and helium ionised regions essentially overlap.  An increase in the abundance
to 0.12 would cause an increase in the theoretical ratio of 20\%, which is
4$\sigma$ above the observed ratio, but it seems unlikely that G45.12+0.13 has
this abundance (Herter et al.\ 1981).  We therefore set a lower bound to the
stellar effective temperature for the exciting star(s) in G45.12+0.13 of
$\sim38000$K.  This is consistent with the upper limit derived above from the
$2^1$P--$2^1$S/Br$\gamma$ ratio.  Our result is in disagreement with the
inferred HeI/HI ratio derived from radio recombination lines by Wink et al.\
(1982), which has been used previously in modelling this source.  It seems
likely that at the densities known to exist in this region both line broadening
and maser effects have a significant impact on the measured RRL line strength.

\subsection{Forbidden Iron Lines}
Since Fe$^+$ has an ionisation potential of only 16.18eV, it is likely to exist
only near the edges of the nebula.  The [FeII] emitting zone will coincide both
with the H and H$^+$ (and S and S$^+$) regions, but sits outside the He$^+$,
O$^{++}$ and S$^{++}$ zones (all species that have been previously observed
either in this work or by Colgan et al.\ 1991).  Since most of the lines that
can be seen have critical densities larger than the peak core density observed,
we adopted a simple model in trying to interpret the [FeII] lines and assumed
that they all arose in the same isothermal, isodensity region.  For the
temperature we have taken the nominal value of T$_e=$7500K  (since
electron temperature does not change radically with position in most nebular
models). In practice, the variation of most [FeII] line ratios discussed here
with T$_e$ is small in any event, since the separation of the first 13 energy
levels is small.  The near infrared [FeII] ratios are primarily indicators of
electron density.

Collision strengths for [FeII] were updated recently by by Pradhan \& Zhang
(1993) and we use this source in what follows.  Transition probabilities were
taken from Nussbaumer \& Storey (1988).  The wavelengths are calculated from
the energy level data of Johansson (1978).  Table 5 gives the ratios of the
observed transitions from a${}^4$D--a${}^6$D and a${}^4$D--a${}^4$F against the
1.257$\mu$m a${}^4$D$_{7/2}-$a${}^6$D$_{9/2}$ line. Also given there are the
expected theoretical ratios. Of these, the easiest to consider are those
transitions that also arise from a${}^4$D$_{7/2}$.  The observable transitions
fall at 1.321, 1.372 and 1.644$\mu$m.  The 1.644$\mu$m line is completely
unresolved from HI Br12, and the flux was derived by fixing Br12 relative to
Br11 and Br13 using recombination theory, and attributing the residual to the
[FeII] line.  With this partition the line strength agrees with the theory.
This also shows that the [FeII] lines give the same extinction as derived from
the HI lines in Section 5.1 above.  The other two line ratios are also
consistent with the theory.

Use of the other line ratios allow us to predict electron densities, and these
are also given in Table 5.  Again blended lines are likely to give the largest
errors.  This is certainly true for the 1.797 and 1.800$\mu$m lines which  are
also weak detections.   The 1.533$\mu$m transition is also clearly visible,
blended with Br18, and is of roughly the correct strength (predicted to be
$\sim4\times10^{-17}$Wm$^{-2}$), if we scale Br18 to Br19 and Br17 as we did
above for Br12.  Lastly, the 1.745$\mu$m transition is blended with the
$7^3$P--$4^3$S HeI transition.  We therefore discount it in this analysis:  it
is worth noting however, that the maximum contribution from both HeI and [FeII]
lines cannot explain the observed flux alone if the atomic data is correct.

The remaining lines can all be fit by gas of moderately high density,
$n_e\sim10^4$cm$^{-3}$. Colgan et al.\ find a core-halo density distribution
fits the radio map under the assumption of spherical symmetry. Their halo
density is compatible with our derived [FeII] density, which is in line with
the expectation that the Fe$^+$ emitting region should be well away from the
exciting source.  Lastly, we have estimated expected fluxes for this density
for all those lines not seen in our spectra, and find that they are below our
detection threshold as expected.

For the four tabulated [FeIII] transitions we used the collision strengths from
Berrington et al.\ (1991) and the transition probabilities from Garstang
(1957).  From estimates of the expected fluxes it is clear that any other
[FeIII] lines in this wavelength region would be unobservable.  The observed
flux ratios fit the model well if we assume a high density ($n_e>
10^4$cm$^{-3}$), except 
for the 2.3485$\mu$m a$^3$G$_5$--a$^3$H$_5$ line which is too strong.
 It is likely that this line
has substantial underlying emission from the higher members of the hydrogen
Pfund series, especially Pf29.  Using similar methods to those described above,
we can fix the flux in the Pf29 line, and find that it contributes $\sim50$\%
of the observed flux.  The resultant expected strength of the [FeIII] line is
then in agreement with the model.  The errors on the data however preclude
the use of this ratio as an accurate measure of the density but the lower
limit derived is consistent with the expected maximum electron density
from the radio data.

Unfortunately it is not possible to use forbidden iron lines as probes of the
ionisation structure of the nebula since the next stage, Fe$^{3+}$, does not
have transitions in the near infrared, and the stage after that has an
ionisation potential higher than He$^{2+}$ which is not observed in this
nebula.  It is likely that the dominant ionisation stage in the nebula is
Fe$^{3+}$ as seen in the model HII region of Oliva, Moorwood \& Danziger
(1989).

\subsection{Forbidden Sulphur Lines} 
The [SII] lines of 1.0289$\mu$m ${}^2P_{3/2}-{}^2D_{3/2}$, 1.0323$\mu$m
${}^2P_{3/2}-{}^2D_{5/2}$, 1.0339$\mu$m ${}^2P_{1/2}-{}^2D_{3/2}$ and
1.0373$\mu$m ${}^2P_{1/2}-{}^2D_{5/2}$ are highly blended, and it is therefore
difficult to estimate their relative line fluxes accurately.  Consideration of
the atomic data (taken from Mendoza 1983)  however, shows that the ratios
between the lines should be nearly constant except when
n$_e>5\times10^4$cm$^{-3}$.  Unfortunately, the individual components are
completely unresolved in our CGS4 data, and are of low signal-to-noise in our
slightly higher resolution IRIS data.  
We therefore opted to  fit the CGS4
data alone 
by considering a simultaneous fit to all four lines, with a single line
width, and the separations determined by the values above.
The results of this fit are given in Table
3.  We should note however that small changes in the fit parameters produced
little difference in the `goodness' of the final fit, but moderately large
changes in the relative line fluxes in the blend.  In particular, the strengths
of the two closest lines at 1.0323 and 1.0339$\mu$m are somewhat arbitrary,
although the total overall flux in these lines together is correct.
The theoretical data shows that the 
ratios of the lines within this blend are all expected to be 
approximately constant
(except at very high density for those arising from different upper levels).
Comparison of this  data with the observations
shows reasonable agreement, within the
large quoted errors, for all ratios except those involving the 1.0323$\mu$m
line.  As noted, the strength of this line is somewhat arbitrary, so 
we conclude that the theory and observations are in substantial agreement.

We also detect the far-red [SIII] 9535\AA\ ${}^1D_2-{}^3P_2$ and 9072\AA\
${}^1D_2-{}^3P_1$ lines in our IRIS data. The observed, dereddened line ratio
is 2.5$\pm$0.4. The theoretical value is 2.48 (again using data from Mendoza
1983) in excellent agreement with this.  Since we see no other features in our
far red IRIS spectrum in the range 0.86$\mu$m--1$\mu$m, we conclude that the
far-red image of Hefele, Wacker \& Weinberger (1977) is entirely due to these
two lines.  This is also consistent with an optical spectrum we have obtained
of the 0.76$\mu$m--1$\mu$m region, which shows that the [SIII] line emission
has considerable extent along the slit, in agreement with the extended $I$-band
object seen by Hefele et al.  Our conclusions in this regard are different to
those of Hoare et al.\ (1991): they estimated the contribution to the $I$-band
image as being due to nebular continuum, Paschen series HI lines and the two
[SIII] lines.  The fact that neither of the first two are in fact seen explains
why they measure a lower inferred A$_v$ from this procedure.  Lastly, we note
that our extinction corrected 9535\AA\ line intensity is
$\sim1.1\times10^{-12}$Wm$^{-2}$ 
in a 1.5$\times$3.2arcsec$^2$ aperture, compared
with that of Lester (1979) as reported by Simpson \& Rubin (1984) of
1.2$\times10^{-12}$Wm$^{-2}$ in a 30arcsec circular aperture.  Given that we
see the [SIII] emission as extended, 
there is clearly a discrepancy between these two values.  For
comparison, the ratios of HI lines between our IRIS and CGS4 data are
$\sim1.7$.  We therefore expect the total [SIII] emergent flux from this HII
region is higher than that measured by Lester.   Assuming the same  scaling
holds for [SIII] as for HI (eg.\ Rubin et al.\ 1994), then we might expect the
total  9535\AA\ flux in the CGS4 beam to be $\sim1.9\times10^{-12}$Wm$^{-2}$.
We use this scaling to allow us to compare the [SIII] fluxes with the CGS4
data in section  6, and also in scaling the [SIII] lines for presentation in
Figure 2.

Colgan et al.\ (1991) and Simpson \& Rubin (1984) have previously presented
fluxes from the mid-infrared [SIII]  lines.  It is worth examining their
conclusions from what we have determined here.   They derive an average
electron density of 6200$\pm$2800cm$^{-3}$ from the 18.7$\mu$m
${}^3P_2-{}^3P_1$ and 33.5$\mu$m ${}^3P_1-{}^3P_0$ lines.  This is inconsistent
with the densities derived by comparing the 9535\AA\ line with these
transitions (both of those give $n_e>10^4$cm$^{-3}$).  There is an alternative
measurement of the 18.7$\mu$m flux  from IRAS LRS data.  Simpson \& Rubin
(1990) find a value which is $\sim$50\% larger than quoted by  Herter et al.,
though using very low signal-to-noise data.  If we accept that this is a limit
to the true flux, then it agrees with the Herter et al.\ value.
However, it is clear that both the
mid-IR [SIII] lines cannot be correct if we accept that our data and extinction
correction are accurate.



\subsection{Other Features}
We have tentatively identified the line at 1.189$\mu$m with [FeII]
a$^2$G$_{7/2}$--a$^4$D$_{7/2}$ (which has a vacuum wavelength of
1.18847$\mu$m).  To obtain an accurate wavelength for this line we made a
simultaneous fit to both this feature and the neighbouring HeI
5${}^3$D--3${}^3$P line, and corrected the first wavelength according to the
error we found for the HeI line.  This gave the `true' wavelength as
$1.1884\pm0.0002\mu$m.  This is coincident with the expected wavelength of the
[FeII] line.  Previous studies in this waveband range have often identified a
feature at this location as due to 1.1886$\mu$m [PII] (eg.\ Rudy et al.\ 
1992).  We cannot formally exclude this line as the correct identification on
the basis of the observed wavelength therefore, since it is just consistent
with the derived error.  The [PII] line has a counterpart at 1.1471$\mu$m, with
an intrinsic strength which is 2.6 times less (though Rudy et al.\ derive an
observed ratio of 2.1).  After including reddening, the observed line would be
$\sim$25\% of the strength of the 1.1886$\mu$m line, which should be marginally
detectable:  no such line is visible however. We note that in at least one
other object, the emission line star $\eta$ Carinae, the same [FeII] line has
been identified (Hamann et al.\ 1994).

The only other features with positive detections in our spectra are those of OI
and H$_2$.  The OI features seen have been discussed previously by Grandi
(1980).  The most likely cause of such strong permitted
oxygen lines is fluorescence.  Given the strength of the 4$^3$S$_0$--3$^3$P
line compared to the 3$^3$D$_0$--3$^3$P line ($\sim$2.1 times stronger), it is
most likely that these features are predominantly excited by UV continuum
radiation.  Similar situations have also been noted in many planetary nebulae
(eg.\ Rudy et al.\ 1992).

All of the lines of molecular hydrogen, H$_2$, seen in our spectra are blended
with other features, with the possible exception of the line seen at
2.285$\mu$m.  The H$_2$ lines are weak in all cases. The relative strengths of
the lines appear to agree well with a model in which the H$_2$ is excited
through UV fluorescence from the hot central star, and in which the ambient
cloud density is high, $n_{\rm H}>10^5$cm$^{-3}$ (Sternberg \& Dalgarno, 1989:
cf the model displayed in their figure 7d).  The inferred
molecular cloud density 
 is consistent with the results of the C$^{34}$S observations of hot
molecular gas near the core of the region (Cesaroni et al.\ 1991).  Given the
low signal-to-noise in the H$_2$ lines it is not feasible to use the line
ratios to do more than assert general agreement with this model however.  The
one anomalous feature is the strength of the 2.285$\mu$m line.  It is possible
that this line is not due to molecular hydrogen, given that it cannot be
explained by the models when the presence of 3--2S(2) emission would imply the
presence of other features in our spectra which are not detected.  Geballe,
Burton \& Isaacman (1991) reached a similar conclusion from a study of
planetary nebulae.  We note that the second unidentified line they discuss at
2.199$\mu$m may also be present in our spectrum, as a weak feature is evident
at 2.2$\mu$m.  Unfortunately that feature is also coincident with 3--2S(3)
H$_2$ emission, and our wavelength calibration in this region is insufficiently
accurate to truly determine the nature of these lines.  The excitation
conditions observed in G45.12+0.13 are however very similar to those described
by Geballe et al.\ for the planetary nebulae in which their unidentified
features occur.  They attribute these features to unidentified fine structure
lines arising from an ion with ionisation potential similar to S$^{3+}$,
which may be weakly detected in G45.12+0.13 (Herter et al.\ 1981, Simpson \&
Rubin 1990, Hoare et al.\ 1991).

\subsection{Near Infrared Continuum}
The major sources of continuum emission in the near infrared are bound-free and
free-free emission from the electrons and ions present (H$^+$ and He$^+$),  the
stellar continuum itself, and emission from both hot dust and possibly  very
small grains out of thermal equilibrium.   Figure 2 has been scaled to show the
observed continuum as well as the emission features.
The presence of
the HI continuum is evident from the obvious Brackett and Pfund discontinuities
(at 1.46 and 2.28$\mu$m respectively). We can quantitatively estimate the
contribution that the HI and HeI continua make to this by calculating the
expected equivalent width of a suitable HI line compared to the continuum.  In
practice we used Pa$\beta$ since there is unlikely to be dust emission at
1.3$\mu$m. We modelled the expected bound-free and free-free components in the
fashion outlined by Brown \& Mathews (1970).  The asymptotic expansions for the
Gaunt factors given by Seaton (1960) were used.  These values agreed to within
1\% with the strictly more correct values tabulated by Ferland (1980) at common
wavelengths between 0.8 and 2.5$\mu$m.   We assumed the HeI continuum was well
approximated by scaling 
the HI continuum, rather than calculating it separately.  This
gives the Pa$\beta$ equivalent width as 0.15$\mu$m when T$_e=7500$K (as usual
there is little sensitivity to density in these calculations).  The measured
equivalent width is  $0.09\pm0.01\mu$m.  In order to match this width with the HI
and HeI continua alone the electron temperature needs to be unfeasibly large
(T$_e\sim20000$K). We also plot in Figure 2
the expected HI and HeI continua for T$_e=8000$K for comparison.   
Figure 4 shows the
continuum after this free-free component has been subtracted and the residual
corrected for extinction. The excess flux near the two HI discontinuities
represents the higher series lines that are completely blended together: the
fact that the continuum on each side of these `features' is essentially flat
shows that the electron temperature we used is a good match for the object (the
best match in terms of this criterion  gives T$_e=8000\rightarrow9000$K in good
agreement with that derived by Wink et al.\ 1982).  

Finally, we have modelled the residual continuum by the mixture of a  crude
stellar continuum (assumed to follow a $\lambda^{-4}$ law in this wavelength
region) and emission from hot dust, assuming a grey-body law for the emissivity
with a power law exponent of one (solid lines in figure 4).  
Clearly both of these components are
required since the residual continuum rises both to the blue and the red. 
Rather than attempt to find a best fit model (since we have  already made
several assumptions), our aim here is to show that this mixture is a good match
to the residual continuum.  The dashed line in figure 4 shows the
 combination of a
power law normalised to fit the continuum at 1.2$\mu$m, and a grey-body
normalised to fit the continuum at 2.4$\mu$m. The dust temperature used was
750K:  other temperatures  produce deviations from the data at 2$\mu$m and
beyond.  However, we note that this temperature itself is sensitive to our
assumption of the form of the emissivity for the dust, and in particular the
power law exponent used therein.  Lastly, we note that despite the fact that we
can see significant stellar continuum emission in our spectra, there is no
evidence for individual stars in the image.  It seems likely that the starlight
is reflected, most likely by dust within the nebula:  only polarimetry
would truly prove this however.

\section{Modelling the emission from G45.12+0.13}
Detailed models of G45.12+0.13 have been presented previously by Simpson \&
Rubin (1984), Colgan et al.\ (1991) and Hoare et al.\ (1991).  The first of
these is largely superceded by Colgan et al., who use the same model, but
better observational data.  However, our data is also superior to the Colgan et
al.\ and Hoare et al.\ models in several respects: first, we have more accurate
measures of the HeI/HI ratio than previously  used; secondly, we have been able
to relax some of the constraints on the [SIII] lines;  thirdly, we have
used both the observed total luminosity and number of ionising photons as a
constraint (Colgan et al.\ do not attempt this explicitly);
lastly, since we have an accurate
measure of the near-IR extinction we can correct these lines properly for the
observed reddening.

We used Cloudy version 84.12 (Ferland 1993) to produce a simple photoionisation
model with suitable parameters for G45.12+0.13.  We used the derived  electron
density profile of Colgan et al.\ since our  Br$\gamma$ image is not
calibrated.  Colgan et al.\ used a distance to the source of  8.1kpc in their
analysis.  Available estimates vary from 6.9kpc (Churchwell, Walmlsley \& Wood
1990) to 9.5kpc (Wink et al.\ 1982).  We will use the distance adopted by
Colgan et al.\ in what follows. We note that if the source were as close as
6.9kpc, then only two O6 stars or one O5 star are required to provide the
necessary observed far-IR luminosity.  The latter can be ruled out immediately
since a star with T$_{eff}\sim45000$K will produce a significant S$^{3+}$
volume, and the observed mid-IR [NeII]/[SIV] ratio indicates this is not true.
We therefore prefer to model the exciting source as a cluster of OB stars, with
a mass function as given by Scalo (1986), in which the number of stars per unit
mass present scales as $M^{-2.85}$, and use the relationship between bolometric
luminosity and mass as calculated by Puxley (1988) from Kurucz (1979) model
atmospheres (Cloudy itself uses somewhat more recent Kurucz models).  The use of
Kurucz models may be problematical in that these models assume line blanketed
LTE is a good approximation for calculating the emergent stellar flux, whereas
we know that the non-LTE line blanketed calculations are actually required. 
However, the latter are extremely difficult to calculate and none are widely
available.

The key observational constraints that any model has to match can be summarised
as follows:  from our data the ratios of the different stages of iron and
sulphur ionisation, and the requirement that helium be fully ionised; from
previous mid-IR data, the ratio of the 10.5$\mu$m [SIV] to  12.8$\mu$m [NeII]
lines which is a sensitive measure of the stellar effective temperature in this
range (this data being taken from the  values quoted by Hoare et al.\ (1991)
from unpublished spectra of Roche  \& Aitken); from far-IR data the strengths
of the 33.5$\mu$m [SIII], 36.0$\mu$m [NeIII], 51.8$\mu$m and 88.4$\mu$m [OIII]
lines (all from Colgan et al).  From Herter et al.\ we convert their 2$\sigma$
detection of [ArII] at 6.99$\mu$m into a 3$\sigma$ limit, and adopt their flux
for the  18.7$\mu$m [SIII] line. We also take as an upper limit the flux of the
15.5$\mu$m [NeIII] line as derived from IRAS LRS data (we derived this limit
from the LRS  data ourselves, rather than take it from Simpson \& Rubin). We
consider a reasonable fit any model line which is within a factor of 2 of the 
observed data, since there are likely to be deviations from the model in any
event introduced by our assumption of spherical symmetry. Lastly, the model
must also have the correct total luminosity and the correct number of ionising
photons (which is equivalent to requiring the actual model electron
distribution be the same as the input data from Colgan et al). These parameters
are taken to be  the total far infrared luminosity (166$\times10^4$\Lsolar) and
the number of ionising photons ($10^{49.53}$s$^{-1}$) required  to produce the
observed radio flux as quoted by Wood \& Churchwell.  By definition, this
also ensures that we match the observed radio flux density from optically thin
free-free emission.

We assumed standard solar abundances since there is little evidence for
deviation from these (e.g. Herter et al).  We did not use this as a free
parameter as Colgan et al.\ did.  We adopt the same extinction curve as Colgan
et al.\ for the mid and far-IR data (see also Simpson \& Rubin 1990).  The
actual values used are given in Table 6. To compare our CGS4 data with the mid
and far-IR data we use the ratio of the Herter et al.\ Br$\gamma$ flux with our
own (1.1): the ratios with Pa$\beta$ given in Table 6 take this factor into
account.  Lastly, we included a typical grain population (see Ferland 1993 for
more details), but with a dust to gas ratio one tenth of the interstellar
value, as found by Hoare et al.\ to be the best fit to their mm data.

The requirement that we match both the number of ionising photons and the total
luminosity is extremely sensitive to the upper and lower mass cut-off we impose
on the initial mass function.   The number of ionising photons is relatively
insensitive to the lower mass limit assumed as long as this is somewhat less
than 20\Msolar ($\sim$30000K).   It is highly sensitive to the upper mass limit
since the the product of the mass function and the expected number of ionising
photons from a star of a given mass is roughly constant above 20\Msolar. 
However, the total luminosity is highly dependent on the lower mass cutoff.  In
practice, unless we truncate the mass function at $\sim20$\Msolar, there are
insufficient hard UV photons to provide a spectrum similar to that observed,
whilst still remaining within the observed total far-IR luminosity.   The
possibility that such a cut-off may exist in reality is hard to gauge. 
Carpenter et al.\ (1993) showed that there is considerable evidence from
infrared imaging that massive stars form in clusters that also contain low mass
stars.  However, Rayner et al.\ (1991) find that although this is true in
general, low luminosity clusters of stars are less common near very massive
stars.  The example they cite of W3A has a cluster of OB stars similar to that
which we find below to give the best fit  for G45.12+0.13.  Therefore, we adopt
this lower mass cut-off for our modelling, and  all the models we describe
below have essentially this form.  

We tried several different models to find a good fit to the data. We varied the
upper mass cut-off between stars with effective temperatures of 38000K and
42000K, we allowed for the `clumped' density model of Colgan et al., and we
also included a lower density `core' to the HII region to try and match the
derived electron density from the [OIII] lines, which Colgan et al.\ found to
be $\sim3000$cm$^{-3}$.  Increasing the upper mass limit beyond this range made
all the line ratios discrepant.  Decreasing the limit below 38000K also made
the ratios discrepant.  The closest fit to the data that we could find had 
a cluster in which the hottest star present
had an effective temperature of 42000K (or a star of $\sim$50\Msolar),
with no clumping and with no low density core.  
The results from this model are given in Table 6. The
major discrepancies we find between this model and the data are as follows:  we
cannot reproduce the observed neon forbidden lines with any degree of accuracy,
with the predicted strengths
of both the 12.8$\mu$m Ne$^+$ line and the 15.4$\mu$m 
Ne$^{++}$ line being too great; we
cannot produce the correct ratio for the [OIII] lines (not surprising since 
this model only has low densities at large radius from the stars), leading 
to the predicted flux of the 88.4$\mu$m line being too small;  
as noted in section 5.4, we cannot match all three of the [SIII] features, with
the predicted strength of the 33.5$\mu$m line being half the observed value;
the  model
also overpredicts the rate at which the 2$^3$S level is photoionised,
consequently making the 2.058$\mu$m line far too weak; lastly, the low
ionisation species ([FeII], [SII]) are predicted to be weaker than seen.
Correctly accounted for are the strengths
of the other sulphur and argon lines, 
the opacity in the helium triplets, and the
helium line strengths other than the 2.058$\mu$m line and the strength of the
stronger [OIII] line and the [NIII] line.

Most of the discrepancies in the model can be set aside.  Some are a
consequences  of our initial assumptions, such as the ratio of the [OIII]
lines.  The weak predicted strength of the 2.058$\mu$m line can be explained by
the HeI 2$^3$S level being predominantly depolulated by collisions to the
singlet levels and not photoionisation, as the model of Clegg \& Harrington
(1989) also shows.  In addition, the predicted strength of the 2.058$\mu$m is
highly dependent on both the dust to gas ratio and on the nature of the dust
grains present.  Since these factors have only a relatively weak impact on the
other lines we have considered it is possible that some of  the difference
between the model and the observations may also lie here. For the discrepancies
with the low ionisation species observed, both [SII] and [FeII] can exist  in
extended partially ionised regions  which were not explicitly allowed for in
the model, and therefore they can only be stronger than the model predicts. The
failure to find a good match to the data when we place a low density region
near the exciting stars (and hence why we cannot reproduce the [OIII] ratios)
probably indicates that spherical symmetry is a poor assumption as noted by
Colgan et al. The major concern is that we cannot achieve a good fit to the
neon lines.  Higher stellar effective temperatures would result in lower
predicted [NeII] flux, but the [SIV] flux would then be too large.  Colgan et
al.\ find a somewhat similar problem in their model, since when they scale the
neon abundance to make the [NeII] line fit, the [NeIII] lines are discrepant. 
We are confident however that the generally good agreement with the model for
most species indicates that the correct model for the stellar population of
G45.12+0.13 is that with a cutoff at both low and high masses such that there
are no stars less massive than $\sim$20\Msolar\  or more massive than
$\sim$50\Msolar.

Lastly, we note that Hoare et al.\ (1991) carried out a similar modelling
process for the mid-IR lines alone.  They quote results only for
two pairs of lines (10.5$\mu$m [SIV]/18.7$\mu$m [SIII] and 15.4$\mu$m 
[NeIII]/12.8$\mu$m [NeII]).  
Only one of their calculations (that using Kurucz atmospheres to model
a cluster of stars) is similar to ours.  However, they do not obtain a good
match to the observed data from their model (they predict ratios of 2.6 and 4.6
respectively).  For the same pairs of lines we predict ratios of 1.0 and 1.4
respectively.  As noted before the neon lines are a poor fit to the observed
data, but our predictions for both pairs are in considerably better agreement
with the observed data than those of Hoare et al.  The major difference
in the assumptions between the models is the density distribution.  Hoare et
al.\ assumed a constant (and relatively low) density.  This largely explains
the difference since the lower density allows the higher energy photons
to penetrate to a larger depth, creating larger volumes of high excitation
states for all ions.  

\section{Conclusions}

We have presented high signal to noise  spectral data covering 0.9--2.5$\mu$m
for the  compact HII region G45.12+0.13.   All of the lines detected by us can
be explained by normal recombination and collisional excitation processes. 
There is no evidence for hot shocked gas (eg.\ Van Buren et al.\ 1990)
indicative of the bow-shock model, nor for the presence of CO emission features
that may have indicated the presence of a remnant accretion disk (Hollenbach et
al.\ 1993).  

We used the HI lines present in our data to estimate  the near-infrared
extinction law in this source and found a  dependence on wavelength that is
somewhat steeper than normally  observed.  After accounting for this extinction
we were able to show that our data are also consistent with a dense core, with
$n_e > 10^4$cm$^{-3}$.  This value is considerably larger than previous
infrared
studies
(eg.\ Colgan et al.\ 1991) have found.  Most of the emission arises from
within the central 3$''\times3''$ region, though our Br$\gamma$ image reveals
considerable extent to the HII region that was not evident from the radio
continuum map of Wood \& Churchwell (1989).   There is tentative evidence that
the higher ionisation species in our data trace higher electron densities,
typical of a core-halo model for the HII region.  

Using an approximation to the true radiative transfer problem, we showed how
the observed HeI triplet lines indicated considerable opacity in the HeI
$n^3$P--2$^3$S series, consistent with the high derived electron density. 
After consideration of all the possible effects we showed that helium must be
fully ionised within the HII region, and that we could set  both upper and
lower limits to the maximum stellar effective temperature of 38000K and 42000K
respectively.  We show how these limits can be set independently of a full
modelling process, and hope more effort will be put into determining the
accuracy of this method theoretically.

Lastly, from a full photoionisation model,  we showed that we can model most of
the observed data on G45.13+0.12 with an OB star cluster which has a restricted
range of  stellar masses present.  In particular, a model where the upper mass
cut-off is set to 50\Msolar\ and the lower mass cut-off to 20\Msolar\ gives a
reasonable fit to the emission line spectrum with the general exception of the
neon lines. We agree with Colgan et al.\ that the asymmetry evident in the
source will probably explain the other differences we found between the
(spherically symmetric) model and the data.

\section*{Acknowledgements}
We would like to thank Melvin Hoare for many useful discussions on
HII regions, Derck Smits for providing a machine
readable copy of his HeI data and
Ruth Doherty for providing the correction coefficients
for the 2.058$\mu$m HeI line.  Part of this work was carried out whilst
SLL was supported by an SERC postdoctoral fellowship at the University
of Oxford.  SLL would also like to thank ANSTO for their financial support
during some of these observations.

\section*{References}
\begin{refs}
\mnref{Allen D.A., Barton, J.R., Burton, M.G., Davies, H., Farrell, T.J.,
      Gillingham, P.R., Lankshear, A.F., Lindner, P.L., Mayfield, D.J., 
      Meadows, V.S., Schafer, G.E., Shortridge, K.E.A., 
      1993, PASA, 10, 298}
\mnref{Berrington, K.A., Zeippen, C.J., Le Dourneuf, M., Eissner, W., Burke,
      P.G., 1991, J.Phys.B:At.Mol.Opt.Phys., 24, 3467}
\mnref{Brown, R.L.,  Mathews, W.G., 1970, \apj, 160, 939}
\mnref{Carpenter, J.M., Snell, R.L., Schloerb, F.P., Skrutskie, M.F., 
      1993, \apj, 407, 657}
\mnref{Cesaroni R., Walmsley C.M., K\"{o}mpe C., Churchwell E.,
      1991, \aaa, 252, 278}
\mnref{Churchwell, E., Walmsley, C.M., Wood, D.O.S., Steppe, H.,
      1990, Radio Recombination Lines: 25 Years of Investigation, p. 73}
\mnref{Churchwell, E., Walmsley, C.M., Wood, D.O.S., 1990, \aas, 83, 119}
\mnref{Clegg, R.E.S., 1987, \mn, 229, 31P}
\mnref{Clegg, R.E.S., Harrington, J.P., 1989, \mn, 239, 869}
\mnref{Colgan, S.W.J., Simpson, J.P., Rubin, R.H., Erickson, E.F.,
      Haas, M.R., Wolf, J., 1991, \apj, 366, 172}
\mnref{Doherty, R.M., Puxley, P.J., Doyon, R., Brand, P.W.J.L., 1994, \mn,
      266, 497}
\mnref{Doherty, R.M., Puxley, P.J., Lumsden, S.L., Doyon, R., 1995, \mn,
      in press}
\mnref{Doyon, R., Puxley, P.J., Joseph, R.D., 1992, \apj, 397, 117}
\mnref{Dyson, J.E., Williams, R.J.R., Redman, M.P., 1996, \mn, in press}
\mnref{Ferland, G.J., 1980, \pasp, 92, 596}
\mnref{Ferland, G.J., 1993, University of Kentucky, Department of Physics
      and Astronomy Internal Report}
\mnref{Garay, G., Reid, M.J., Moran, J.M., 1985, \apj, 289, 681}
\mnref{Garstang, R.H., 1957, \mn, 117, 393}
\mnref{Geballe, T.R., Burton, M.G., Isaacman, R., 1991, \mn, 253, 75}
\mnref{Grandi, S.A., 1980, \apj, 238, 10}
\mnref{Habing, H.J., Israel, F.P., 1979,  ARA\&A, 17, 345}
\mnref{Hamann, F., DePoy, D.L., Johansson, S., Elias, J., 1994, \apj, 422, 626}
\mnref{Hefele, H., Wacker, W., Weinberger, R., 1977, \aaa, 56, 407}
\mnref{Herter T., Helfer H.L., Pipher J.L., Forrest W.J.,
      McCarthy J., Houck J.R., Willner S.P., Puetter R.C., Rudy R.J.,
      1981, \apj, 250, 186}
\mnref{Hoare, M.G., Roche, P.F., Glencross, W.N., 1991, \mn, 251, 584}
\mnref{Hollenbach, D., Johnstone, D., Shu, F., \apj, 428, 654} 
\mnref{Hummer, D.G., Storey, P.J., 1987, \mn, 224, 801}
\mnref{Johansson, S., 1978, Physica Scripta, 18, 217}
\mnref{Keto, E.R., Ho, P.T.P., Haschick, A.D., 1987, \apj, 318, 712}
\mnref{Kurucz, R.L., 1979, \apjs, 40, 1}
\mnref{Landini, M., Natta, A., Oliva, E., Salinari, P., Moorwood, A.F.M., 
      1984, \aaa, 134, 284}
\mnref{Lester, D.F., 1979, Ph.D. Thesis, University of California,
      Santa Cruz}
\mnref{Lumsden, S.L., Hoare, M.G., 1995, \apj, in press, to appear Jun 20}
\mnref{Lumsden, S.L., Puxley, P.J., Doherty, R.M., 1994, \mn, 268, 821}
\mnref{Matthews, H.E., Goss, W.M., Winnberg, A., Habing, H.J., 1977, \aaa, 61,
      261}
\mnref{Mendoza, C.,  1983, in IAU Symposium 105, Planetary Nebulae,
      ed. D.R. Flower, Reidel, Dordrecht, p143}
\mnref{Mountain, C.M., Robertson, D.J., Lee, T.J., Wade, R., 1990, 
      in Crawford, D.L., ed., Instrumentation in Astronomy VII, 
      Proc. SPIE, 1235, 25}
\mnref{Natta, A., Panagia, N., 1984, \apj, 287, 228}
\mnref{Nussbaumer, H., Storey, P.J., 1988, \aaa, 193, 327}
\mnref{Oliva, E., Moorwood, A.F.M., Danziger, I.J., 1989, \aaa, 214, 307}
\mnref{Osterbrock, D.E., 1989, Astrophysics of Gaseous Nebulae and AGN,
      University Science Books, Mill Valley, USA}
\mnref{Pradhan, A.K., Zhang, H.L., 1993, \apj, 409, L77}
\mnref{Puxley, P.J., 1988, PhD thesis, University of Edinburgh}
\mnref{Robbins, R.R., 1968, \apj, 151, 511}
\mnref{Rayner, J., Hodapp, K., Zinnecker, H., 1991, in Astrophysics with
      Infrared Arrays, ASP Conference Series, Vol 14, p264}
\mnref{Rubin, R.H.,  1968, \apj, 154, 391}
\mnref{Rudy, R.J., Erwin, P., Rossano, G.S., Puetter, R.C., 1992, \apj, 
      384, 536}
\mnref{Sawey, P.M.J., Berrington, K.A., 1993, At. Data \& Nuc. Data Tables,
      55, 81}
\mnref{Scalo, J.M., 1986, Fund. Cosm. Phys., 11, 1}
\mnref{Shields, J.C., 1993, \apj, 419, 181}
\mnref{Simpson, J.P., Rubin, R.H., 1984, \apj, 281, 184}
\mnref{Simpson, J.P., Rubin, R.H., 1990, \apj, 354, 165}
\mnref{Smits, D.P., 1991a, \mn, 248, 193}
\mnref{Smits, D.P., 1991b, \mn, 251, 316}
\mnref{Sternberg, A., Dalgarno, A.,  1989, \apj, 338, 197}
\mnref{Tanaka, M., Yamashita, T., Sato, S., Nislud, M., 1983, \pasp, 97, 1112}
\mnref{Van Buren, D., MacLow, M.-M., Wood, D.O.S., Churchwell, E., 1990,
\apj, 353, 570}
\mnref{Wink, J.E., Altenhoff, W.J., Mezger, P.G., 1982, \aaa, 108, 227}
\mnref{Wood, D.O.S.,  Churchwell, E., 1989, \apjs, 69, 831}
\mnref{Wood, D.O.S.,  Churchwell, E., Salter, C.J., 1988, \apj, 325, 694}
\mnref{Wood, D.O.S., Handa, T., Fukui, Y., Churchwell, E., Sofu, Y., Iwata, T.,
      1988, \apj, 326, 884}
\end{refs}

\newpage\onecolumn
\begin{table}
\begin{tabular}{ccccc}
Date & Wavelength ($\mu$m) & Resolution (R) & Integration Time (s) &
Airmass \\
1991 October 24  &   1--1.2 & 340 & 1600 & 1.64 \\
1991 October 25  & 1.2--1.4 & 340 & 1280 & 1.14 \\
1991 October 25  & 1.4--1.6 & 470 &  960 & 1.59 \\
1992 November 15 &  2--2.4  & 340 &  320 & 1.80 \\
1993 April 4     & 1.6--1.8 & 530 &  960 & 1.53 \\
1993 April 4     & 1.7--2.1 & 340 &  960 & 1.25 \\
1993 April 4     & 2.3--2.5 & 700 &  640 & 1.04 \\
\end{tabular}
\caption{Details of the UKIRT Observations.}
\begin{tabular}{ccccc}
Date & Wavelength ($\mu$m) & Resolution (R) & Integration Time (s) &
Airmass \\
1993 September 23 & 0.86--1.5 & 420 &  1800 & 1.36 \\
1993 September 23 & 1.4--2.5 & 420 &  600 & 1.35 \\
\end{tabular}
\caption{Details of the AAT Observations.}
\end{table}

\begin{table}
\begin{tabular}{lclcrr}
$\lambda_{\rm obs}$ ($\mu$m) & $\delta\lambda$ ($\mu$m) &
      Line I.D. & $\lambda_{\rm line}$ ($\mu$m) & 
      \multicolumn{2}{c}{Corrected line strength} \\
&($\times10^4$) & & & CGS4 & IRIS \\
0.9075 & 2 & [SIII] ${}^1$D$_2-{}^3$P$_1$    & 0.9072 & & $467\pm29$ \\
0.9537 & 2 & [SIII] ${}^1$D$_2-{}^3$P$_2$    & 0.9535 & & $1160\pm120$\\
1.0052 & 3 &  Pa$\delta$                  & 1.0052 & $43\pm3$   & $37\pm6$\\
1.0298$bl$ & 9 & [SII] ${}^2P_{3/2}-{}^2D_{3/2}$ & 1.0289 & $3.0\pm1.0$ &  \\
1.0332$bl$ & 9 & [SII] ${}^2P_{3/2}-{}^2D_{5/2}$ & 1.0323 & $6.0\pm0.8$ & \\
1.0348$bl$ & 9 & [SII] ${}^2P_{1/2}-{}^2D_{3/2}$ & 1.0339 & $3.2\pm1.7$ & \\
1.0382$bl$ & 9 & [SII] ${}^2P_{1/2}-{}^2D_{5/2}$ & 1.0373 & $1.2\pm0.5$ & \\
1.0837 & 3 & HeI, 2$^3$P--2$^3$S  & 1.0833 & $495\pm19$ & $497\pm11$\\
1.0944 & 3 &  Pa$\gamma$        & 1.0941 & $64\pm2$  & $63\pm2$ \\
1.1296 & 3 & OI, 3$^3$D$_0$--3$^3$P  & 1.1289  & $1.0\pm0.2$ & \\
1.1887 & 3 & [FeII] a$^2$G$_{7/2}$--a$^4$D$_{7/2}$?  & 1.1885 & $0.7\pm0.1$\\
1.1973 & 3 & HeI, 5$^3$D--3$^3$P             & 1.1972 & $1.6\pm0.2$ & $1.2\pm0.1$\\
1.2534 & 3 & HeI, 4$^3$P--3$^3$S             & 1.2531 & $2.9\pm0.1$  & $2.7\pm0.1$\\
1.2574 & 3 & [FeII] a$^4$D$_{7/2}$--a$^6$D$_{9/2}$ &  1.2570 & $1.7\pm0.1$ & $1.1\pm0.1$ \\
1.2800$bl$ & 3 & HeI, 5$^{3}$F--3$^{3}$D    & 1.2789 & $11.3\pm0.5 bl$ &
$6.9\pm0.8 bl$\\
1.2800$bl$ & 3 & HeI, 5$^{1}$F--3$^{3}$D    & 1.2794 & $11.3\pm0.5 bl$ &
$6.9\pm0.8 bl$\\
1.2827 & 3 &  Pa$\beta$   & 1.2821 & $100\pm$4 & $100\pm$2 \\
1.2941$bl$ & 4 & [FeII] a$^4$D$_{5/2}$--a$^6$D$_{5/2}$ & 1.2946 & 
$0.4\pm0.1 bl$ & \\
1.2980$bl$ & 4 & [FeII] a$^4$D$_{3/2}$--a$^6$D$_{1/2}$ & 1.2981 & 
$0.4\pm0.1 bl$ & \\
1.2984$bl$ & 4 & HeI, 5$^3$P--3$^3$D  & 1.2988 & $1.1\pm0.1$ & $0.9\pm0.2$\\
1.3171$bl$& 4 & OI, 4$^3$S$_0$--3$^3$P & 1.3168   & $2.1\pm0.1$ & $1.5\pm0.2$\\
1.3213$bl$ &  3 & [FeII] a$^4$D$_{7/2}$--a$^6$D$_{7/2}$ & 1.3209  & $0.4\pm0.1$ & $0.3\pm0.1$\\
1.3724 &  6 & [FeII] a$^4$D$_{7/2}$--a$^6$D$_{5/2}$ & 1.3722  & $0.3\pm0.1$ & \\
1.5086$bl$ &   5   &  Br22--4  & 1.5087 & $1.2\pm0.1bl$ & \\
1.5086$bl$ &   5   & HeI, 4$^1$P--3$^1$S & 1.5088  & $1.2\pm0.1bl$ & \\
1.5138 &  3 &  Br21--4   & 1.5137 & $0.9\pm0.1$ & \\
1.5196 &  3 &  Br20--4   & 1.5196 & $1.0\pm0.1$ & \\
1.5266 &  3 &  Br19--4   & 1.5265 & $0.9\pm0.1$ & \\
1.5345$bl$ &  5 &  Br18--4   & 1.5346 & $1.5\pm0.1 bl$ & $1.2\pm0.2 bl$ \\
1.5345$bl$ & 5 & [FeII] a$^4$D$_{5/2}$--a$^4$F$_{9/2}$ & 1.5339 & $1.5\pm0.1
bl$ &$1.2\pm0.2 bl$\\
1.5444 &  3 &  Br17--4   & 1.5443 & $1.5\pm0.1$ & $1.2\pm0.1$\\
1.5562 &  3 &  Br16--4   & 1.5561 & $1.8\pm0.1$ & $1.6\pm0.1$\\
1.5708 &  3 &  Br15--4   & 1.5705 & $2.1\pm0.1$ & $1.6\pm0.1$\\
1.5888 &  3 &  Br14--4   & 1.5885 & $2.6\pm0.2$ & $2.3\pm0.1$\\
1.6005 &  3 & [FeII] a$^4$D$_{3/2}$--a$^4$F$_{7/2}$ ? & 1.5999 & $0.2\pm0.1$ & \\
1.6117 &  3 &  Br13--4   & 1.6114 & $2.7\pm0.1$ & $2.8\pm0.1$\\
1.6414$bl$ &  4 &  Br12--4   & 1.6412 & $3.9\pm0.2$ & $4.2\pm0.1$\\
1.6444$bl$ &  4 & [FeII] a$^4$D$_{7/2}$--a$^4$F$_{9/2}$ & 1.6440 & $3.9\pm0.2$
 & $4.2\pm0.1$\\
1.6640 & 5 & [FeII] a$^4$D$_{1/2}$--a$^4$F$_{5/2}$ & 1.6642 & $0.10\pm0.07$ \\
1.6776 & 3 & [FeII] a$^4$D$_{5/2}$--a$^4$F$_{7/2}$ & 1.6773 & $0.3\pm0.1$ & \\
1.6812 & 3 &  Br11--4    & 1.6811 & $4.4\pm0.2$ & $4.9\pm0.2$\\
\end{tabular}
\caption{Identifications of detected lines.  
Identifications which are unsure are 
marked with a ?. The Pa$\beta$ flux after correction for extinction is
1.64$\times10^{-13}$Wm$^{-2}$ (CGS4) and 9.83$\times10^{-14}$Wm$^{-2}$ (IRIS).
All strengths are quoted relative to $F({\mbox Pa}\beta)=100$.}
\end{table}
\newpage\onecolumn
\addtocounter{table}{-1}
\begin{table}
\begin{tabular}{lclcrr}
$\lambda_{\rm obs}$ ($\mu$m) & $\delta\lambda$ ($\mu$m) &
      Line I.D. & $\lambda_{\rm line}$ ($\mu$m) & 
      \multicolumn{2}{c}{Corrected line strength} \\
&($\times10^4$) & & & CGS4 & IRIS \\
1.7011      &  3 & HeI, 4$^3$D--3$^3$P & 1.7007 & $2.1\pm0.1$ & $2.5\pm0.1$\\
1.7370      &  3 &  Br10--4         & 1.7367 & $6.3\pm0.2$ & $6.8\pm$0.1\\
1.7456$bl$    &  3 & [FeII] a$^4$D$_{1/2}$--a$^4$F$_{3/2}$ & 1.7454 & 
$0.3\pm0.1 bl$& $0.4\pm0.1 bl$\\
1.7456$bl$    &  3 & HeI, 7$^3$P--4$^3$S & 1.7455 & $0.3\pm0.1 bl$& $0.4\pm0.1
bl$\\
1.7978$bl$    &  3 & [FeII] a$^4$D$_{3/2}$--a$^4$F$_{3/2}$ ? & 1.7976 &
$0.2\pm0.1 bl$& \\
1.8008$bl$    &  3 & [FeII] a$^4$D$_{5/2}$--a$^4$F$_{5/2}$ ? & 1.8005 &
$0.3\pm0.1 bl$& \\
1.8178      &  5 &  Br9--4           & 1.8179 & $11.4\pm0.4$ & $6.0\pm0.1$\\
1.8746      &  7 &  Pa$\alpha$       & 1.8756 & $305\pm12$ & \\
1.9093      &  5 & HeI, 4$^1$D--3$^1$P & 1.9095  & $1.2\pm0.1$ & \\
1.9449      &  5 &  Br$\delta$       & 1.9451 & $14.5\pm0.6$ & $14.2\pm0.1$\\
1.9547      &  5 & HeI, 4$^3$P--3$^3$D & 1.9549  & $2.1\pm0.1$ & $1.3\pm0.1$\\
2.0583      &  5 & HeI, 2$^1$P--2$^1$S & 2.0587  & $23.2\pm0.9$& $19.2\pm0.2$\\
2.1128$bl$  &  5 & HeI, 4$^{3}$S--3$^{3}$P & 2.1126 & $1.0\pm0.1 bl$ 
& $1.1\pm0.1 bl$\\
2.1128$bl$  &  5 & HeI, 4$^{1}$S--3$^{1}$P & 2.1138 & $1.0\pm0.1 bl$ & 
$1.1\pm0.1 bl$\\
2.1210$bl$? &  5 & H$_2$, 2--1S(0)     &  2.1218     & $0.09\pm0.06 bl$ & \\
2.1460      &  5 & [FeIII] a$^3$G$_3$--a$^3$H$_4$? & 2.1457& $0.16\pm0.05$  & 
$0.18\pm0.05$\\
2.1657      &  5 &  Br$\gamma$          & 2.1661 & $19.0\pm1.0$ & $18.5\pm0.1$\\
2.2185      &  5 & [FeIII] a$^3$G$_5$--a$^3$H$_6$ & 2.2183 & $0.42\pm0.05$  &
$0.28\pm0.05$\\
2.2418      &  5 & [FeIII] a$^3$G$_4$--a$^3$H$_4$ ?  & 2.2427 & $0.21\pm0.05$ &
$0.16\pm0.05$\\
2.2854      &  5 &  ?   &          & $0.27\pm0.06$ & $0.22\pm0.05$\\
2.3487$bl$    &  3 & [FeIII] a$^3$G$_5$--a$^3$H$_5$ & 2.3485 & $0.44\pm0.06 bl$& \\
2.3487$bl$    &  3 &  Pf29--5     & 2.3492 & $0.44\pm0.06 bl$& \\
2.3662      &  3 &  Pf26--5     & 2.3669 & $0.28\pm0.10$ & \\
2.3737      &  3 &  Pf25--5     & 2.3744 & $0.28\pm0.08$& \\
2.3824      &  3 &  Pf24--5     & 2.3828 & $0.32\pm0.10$& \\
2.3919      &  3 &  Pf23--5     & 2.3925 & $0.33\pm0.06$& \\
2.4033$bl$    &  3 &  Pf22--5     & 2.4036 & $0.40\pm0.07 bl$& \\
2.4063$bl$    &  5 & H$_2$, 1--0 Q$_1$& 2.4066 & $0.16\pm0.07 bl$ \\
2.4155      &  3 &  Pf21--5     & 2.4164 & $0.66\pm0.31$& \\
2.4310      &  3 &  Pf20--5     & 2.4314 & $0.43\pm0.10$& \\
2.4486      &  3 &  Pf19--5     & 2.4490 & $0.73\pm0.11$& \\
2.4693      &  3 &  Pf18--5     & 2.4700 & $0.74\pm0.11$& \\
2.4942      &  3 &  Pf17--5     & 2.4953 & $0.87\pm0.15$& \\
\end{tabular}
\caption{
Identifications of detected lines.  
Identifications which are unsure are 
marked with a ?. 
All strengths are quoted relative to $F({\mbox Pa}\beta)=100$.}
\end{table}

\begin{table}
\begin{tabular}{lcccc}
Line ID & $\lambda(\mu$m) & Observed Ratio & \multicolumn{2}{c}{Theoretical
Ratio (Smits)} \\
 & &  & T$_e=$5000K &  T$_e=$10000K \\
2$^3$P--2$^3$S & 1.083 & 235$\pm$14 & 165 & 450\\
5$^3$D--3$^3$P & 1.198 & 0.76$\pm$0.10 & 0.67 & 0.70 \\
4$^3$P--3$^3$S & 1.253 & 1.38$\pm$0.08 & 0.33 & 0.41 \\
5$^{1,3}$F--3$^{1,3}$D & 1.279$bl$ & 5.4$\pm2.0$ & 3.4 & 3.0 \\
5$^3$P--3$^3$D & 1.298$bl$ & 0.52$\pm$0.05 & 0.11 & 0.13 \\
4$^1$P--3$^1$S & 1.508$bl$ & 0.19$\pm$0.05 & 0.15 & 0.17 \\
7$^3$P--4$^3$S & 1.745$bl$ & 0.16$\pm$0.06 & 0.02 & 0.03 \\
4$^1$D--3$^1$P & 1.909 & 0.57$\pm$0.05 & 0.36 & 0.35 \\ 
4$^3$P--3$^3$D & 1.955 & 1.00$\pm$0.06 & 0.21 & 0.26 \\
2$^1$P--2$^1$S & 2.058 & 11.0$\pm$0.7 & 10.9 & 15.6 \\
4$^{1,3}$S--3$^{1,3}$P & 2.113$bl$ & 0.48$\pm0.05$ & 0.24 & 0.33 \\
\end{tabular}
\caption{Observed and predicted HeI line ratios against the 1.7007$\mu$m
4$^3$D--3$^3$P transition.  Theoretical values are given for
T$_e=5000$K and 10000K, and are independent of density, with the exception of
the 2P-2S lines.  These are quoted for $n_e=10^4$cm$^{-3}$, and are also
further discussed in the text since they are affected by processes not included
in Smits theoretical data.
The observed values come
from the CGS4 data.  A $bl$ after the wavelength indicates a line that is 
completely blended with another feature: the strength of the line
has been derived as indicated in the text.  Only those lines blended with
HI features are likely to have completely accurate line strengths quoted.}
\end{table}

\onecolumn
\begin{table}
\begin{tabular}{ccccl}
Line & $\lambda(\mu$m) & Observed Ratio & Theoretical Ratio & $n_e$ (cm$^{-3}$)\\
a$^4$D$_{5/2}$--a$^6$D$_{5/2}$&1.294 & 0.24$\pm$0.06 & 0.02--0.25 &
10000$^{+30000}_{-6000}$ \\
a$^4$D$_{3/2}$--a$^6$D$_{1/2}$&1.298 & 0.24$\pm$0.06 & 0.003--0.08 &
10000$^{+30000}_{-6000}$ \\
a$^4$D$_{7/2}$--a$^6$D$_{7/2}$&1.321 & 0.24$\pm$0.06 & 0.262 \\
a$^4$D$_{7/2}$--a$^6$D$_{5/2}$&1.372 & 0.16$\pm$0.06 & 0.171 \\
a$^4$D$_{3/2}$--a$^4$F$_{7/2}$&1.601 & 0.14$\pm$0.05 & 0.008--0.2 & 15000$^{+35000}_{-8000}$    \\
a$^4$D$_{7/2}$--a$^4$F$_{9/2}$&1.644 & 0.67$\pm$0.05 & 0.74 \\
a$^4$D$_{1/2}$--a$^4$F$_{5/2}$&1.664 & 0.06$\pm$0.05 & 0.003--0.12 & 10000$^{+50000}_{-9000}$    \\
a$^4$D$_{5/2}$--a$^4$F$_{7/2}$&1.677 & 0.18$\pm$0.06 & 0.01--0.2 & 30000$^{+\infty}_{-20000}$    \\
a$^4$D$_{1/2}$--a$^4$F$_{3/2}$&1.745 & 0.20$\pm$0.04 & 0.002--0.06 \\
a$^4$D$_{3/2}$--a$^4$F$_{3/2}$&1.798 & 0.12$\pm$0.06 & 0.004--0.1 & $>15000$ \\
a$^4$D$_{5/2}$--a$^4$F$_{5/2}$&1.801 & 0.18$\pm$0.06 & 0.009--0.14 & $>15000$ \\
\end{tabular}
\caption{Observed and predicted [FeII] line ratios against the 1.257$\mu$m
a$^6$D$_{9/2}$--a$^4$D$_{7/2}$ transition.  Theoretical values assume
T$_e=7500$K and span the density range 10--10$^6$cm$^{-3}$.  Where
only one value is given, this transition shares a common upper level with the
1.257$\mu$m line.  The pair of lines at 1.294/1.298$\mu$m are blended,
so we have used the sum of the theoretical data to match the observed value.
}
\end{table}

\begin{table}
\begin{tabular}{clcccc}
Line & $\lambda(\mu$m) & Observed Flux  & Dereddened Flux
 & Observed Ratio with Pa$\beta$ & Model Ratio with Pa$\beta$\\
 & & \multicolumn{2}{c}{(10$^{-14}$ Wm$^{-2}$)} \\
\null [SIII] & 0.953 & & & 11.6$\pm1.2$ & 10.0 \\
\null [SII] & 1.03 & & &  0.13$\pm$0.02 & 0.05 \\
\null HeI  & 1.083 & & &  5.0$\pm$0.2 & 4.2 \\
\null [FeII] & 1.257 & & & $>0.03$ & 0.007 \\
\null HeI  & 1.701 & & &  0.02$\pm$0.001 & 0.02 \\    
\null HeI  & 2.058 & & &  0.2$\pm$0.01 & 0.06 \\
\null [ArII] & 6.99 &$<7.8$ & $<$13.5 & $<0.55$ & 0.02 \\
\null [ArIII] &8.99 & 3.3$\pm$0.4 & 20$\pm2$ &0.8$\pm$0.1 & 1.1 \\
\null [SIV] &  10.5& $<12$ & $<80$ & $<3.4$ & 1.5\\
\null [NeII] & 12.8 &26$\pm$2.6 & 52$\pm5$ & 2.2$\pm0.3$ & 3.4\\
\null [NeIII] &15.4 & $<19$ & $<36$ & $<1.5$ & 4.6\\
\null [SIII] & 18.7 &$15.8\pm1.9$ & $38\pm5$ & $1.5\pm0.2$ & 1.5 \\
\null [SIII] & 33.5 &7.4$\pm0.7$ & $10.0\pm1.0$ & 0.42$\pm0.05$ & 0.20 \\
\null [NeIII] &36.0 & 5.0$\pm0.7$ & $6.4\pm0.8$& $0.26\pm0.03$ &0.23\\
\null [OIII] & 51.8 &17.4$\pm1.3$ & $20\pm1.5$ &$0.84\pm0.06$&1.0\\
\null [NIII] & 57.3 &3.6$\pm0.7$ & $4.0\pm0.8$ &$0.16\pm0.03$&0.13\\
\null [OIII] & 88.4 &4.6$\pm0.3$ & $5.0\pm0.5$ &$0.21\pm0.02$&0.11\\
\end{tabular}
\caption{Adopted mid and far-IR line fluxes from the literature.  Sources
are given in the text.  The extinction correction applied follows that
of Colgan et al. The near-IR values are taken from Table 3, and only the ratio
with Pa$\beta$ is repeated here.  The values for
the [FeII] and [SII] lines are summed over the relevant multiplets to match
the output from Cloudy.  Since
not all the the [FeII] lines were observed, a lower limit is given by
summing all those observed.
The model ratios
are for the best fitting photoionisation model as described in the text.}
\end{table}

\clearpage

\hspace*{0in}\begin{minipage}{5.8in}{\begin{center}
\leavevmode
\epsfxsize=5in 
\epsfbox{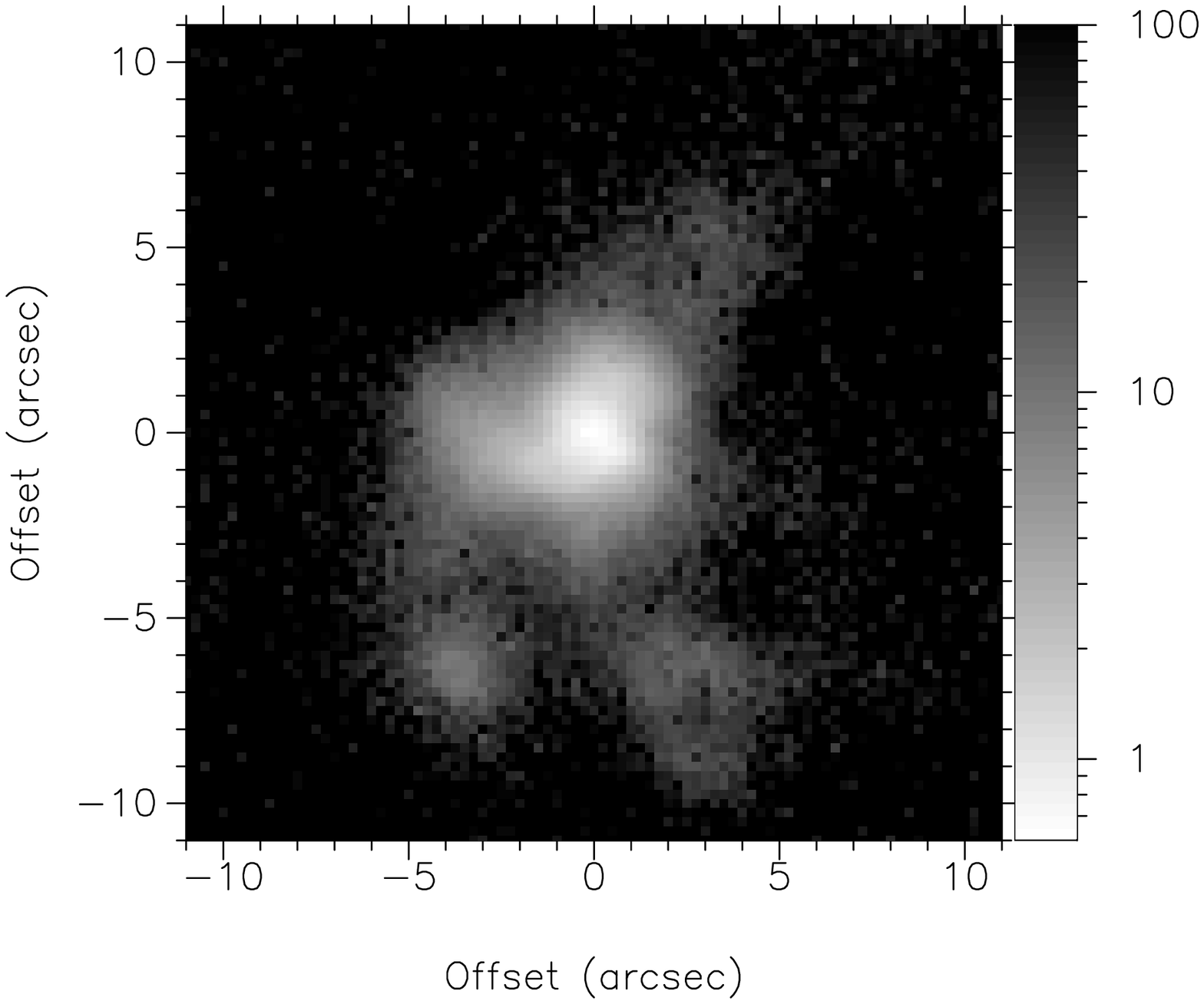}
\end{center}}\end{minipage}

\vspace*{0.1in}
\noindent{\bf Figure 1:}  Greyscale Br$\gamma$ 
image of G45.12+0.13.  The log
of the observed intensity is plotted.  No absolute calibration is given
due to the non-photometric conditions during these observations.
The (0,0) position is assumed to correspond to the peak in the 2cm
radio map.  Extended emission not evident in the radio map (figure
23a of Wood \& Churchwell 1989) is obvious in our map.\\

\hspace*{0in}\begin{minipage}{5.8in}{\begin{center}
\leavevmode
\epsfxsize=5in
\epsfbox{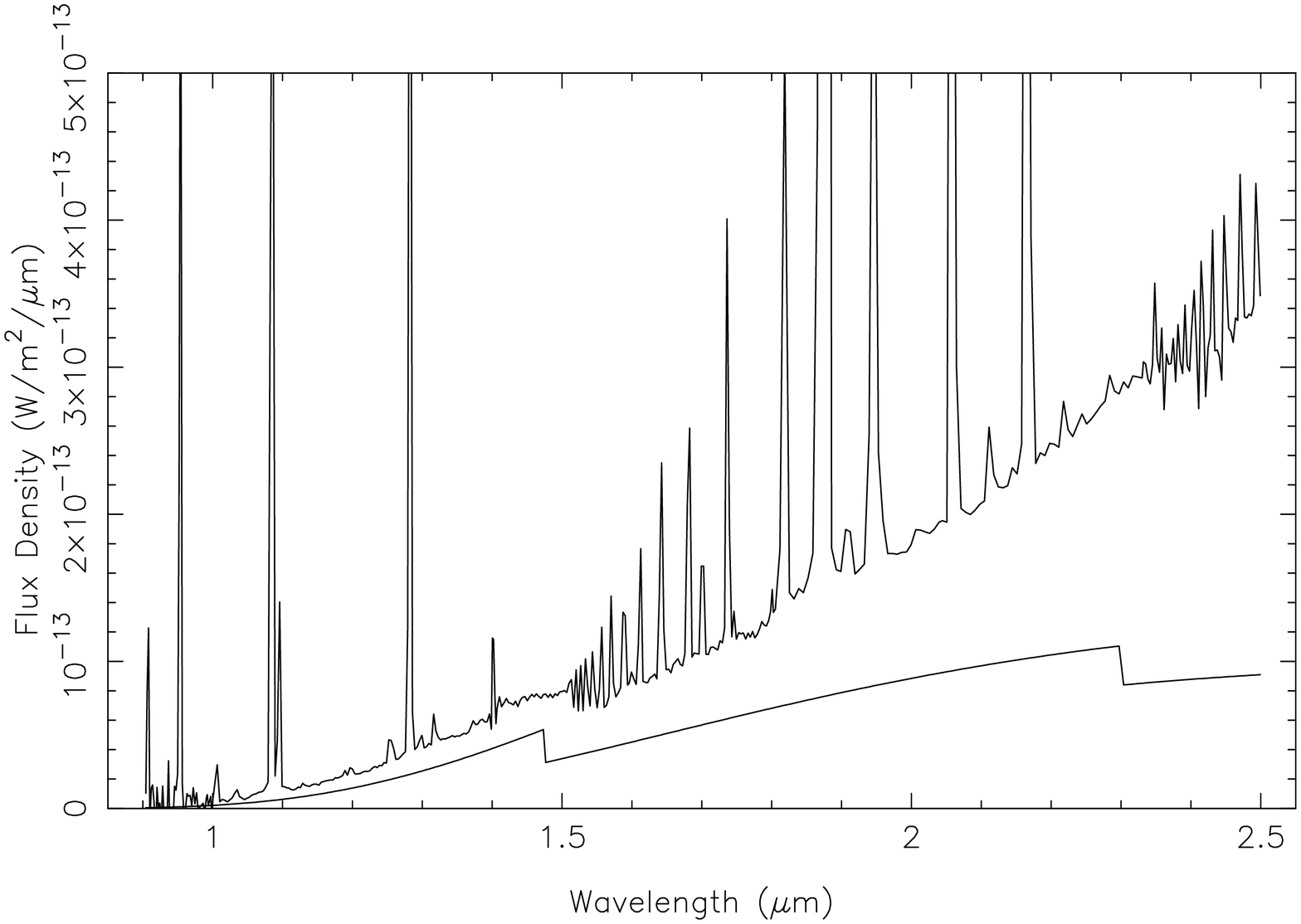}
\end{center}}\end{minipage}

\vspace*{0.1in}
\noindent{\bf Figure 2:}  Observed spectrum of G45.12+0.13 with the
stronger lines suppressed to show the weaker features present.
The solid
line at the bottom is the `best' match to the expected free-free
component (scaled to the observed Pa$\beta$ line strength) as described in
section 5.6, including
the correct allowance for the observed extinction.\\

\hspace*{0in}\begin{minipage}{6in}{\begin{center}
\leavevmode
\epsfxsize=3in
\epsfbox{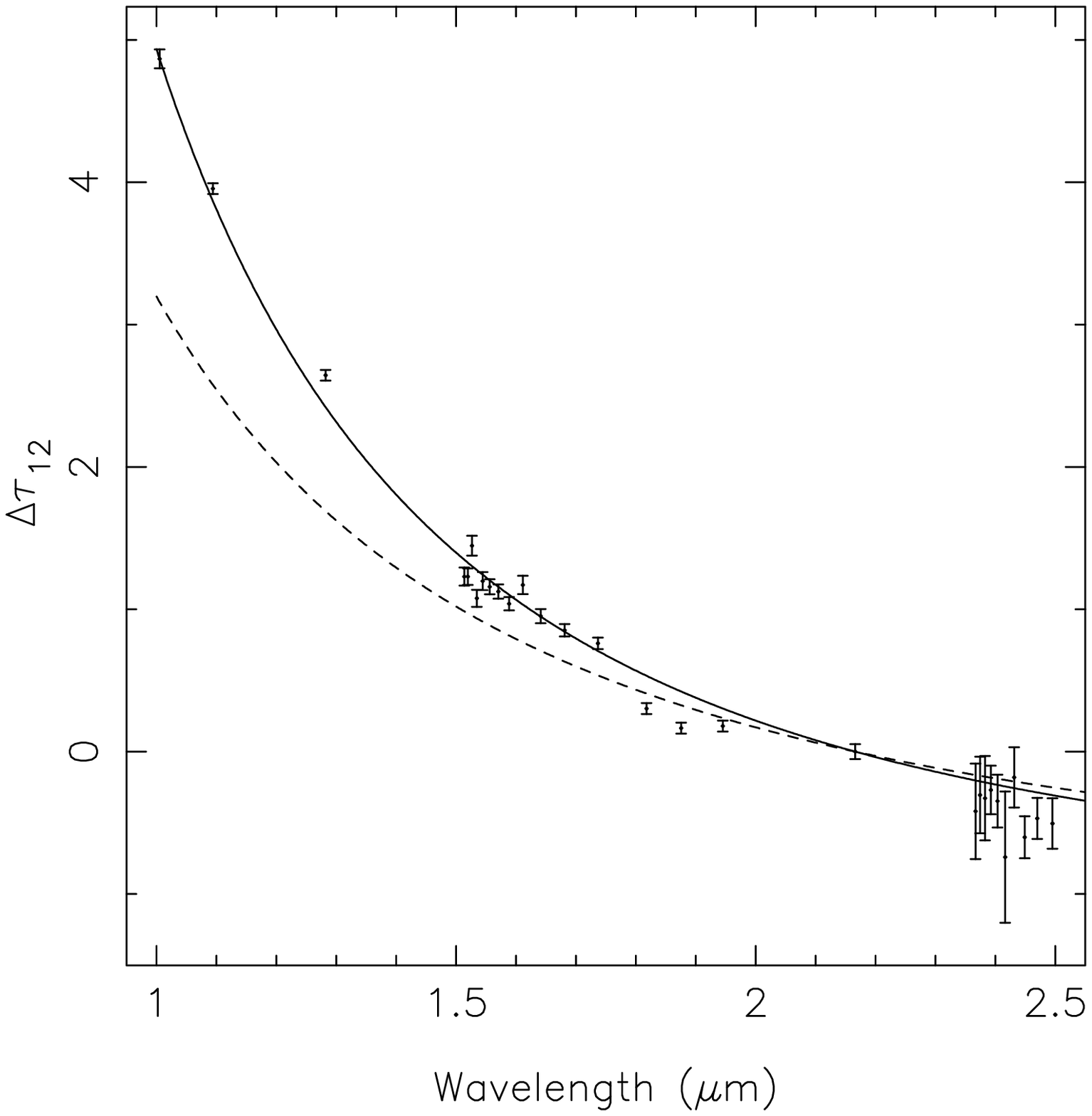}\epsfxsize=3in\epsfbox{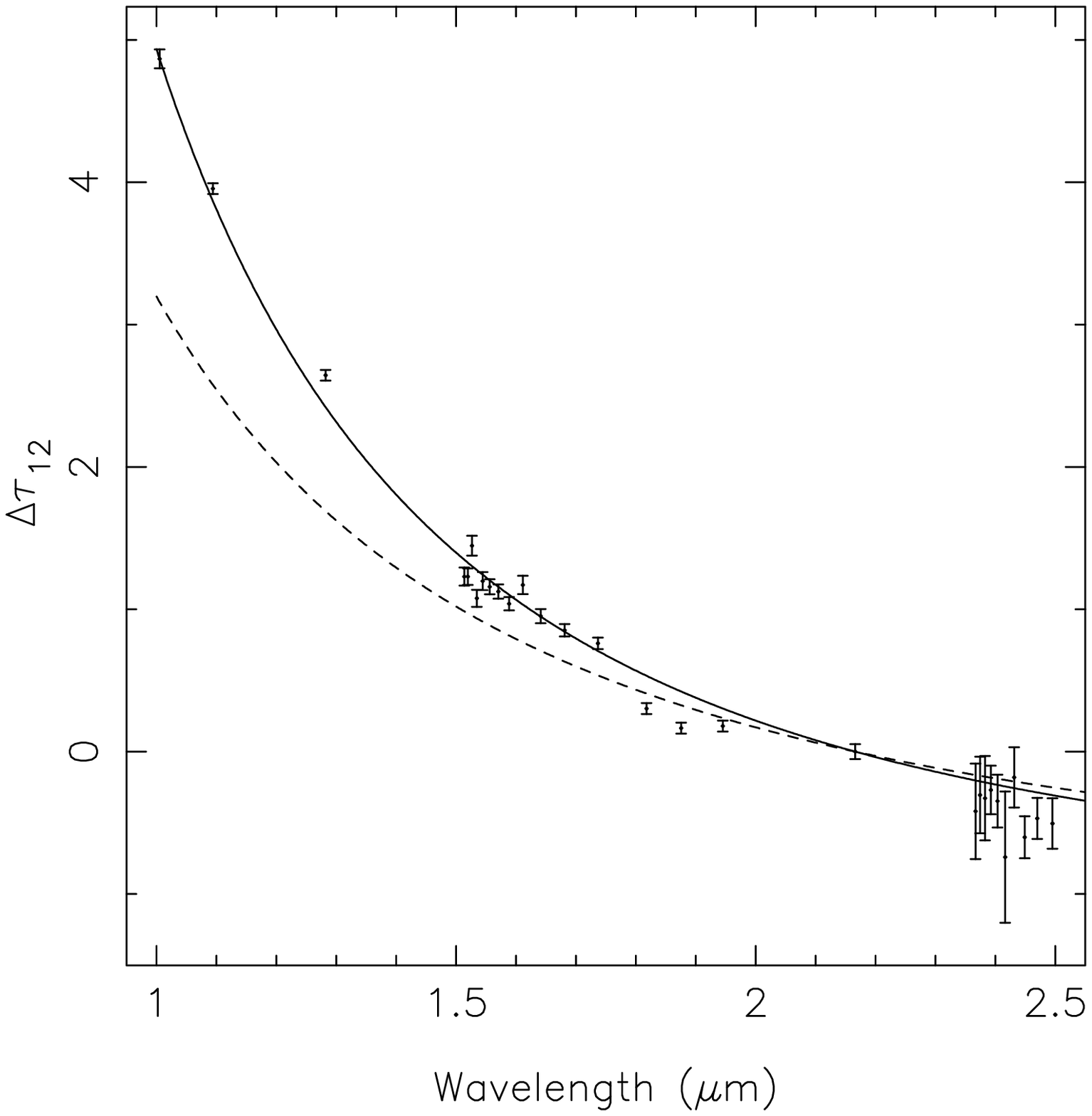}
\end{center}}\end{minipage}

\vspace*{0.1in}
\noindent{\bf Figure 3:}  Observed values of the extinction for (a)
the CGS4 data and (b) the IRIS data.  The derived best fit to $\alpha$ and
$\tau_{{\rm Br}\gamma}$ is shown as a solid line.  The dashed line represents
the extinction law given by Landini et al.\ (1984).\\

\hspace*{0in}\begin{minipage}{5.8in}{\begin{center}
\leavevmode
\epsfxsize=5in
\epsfbox{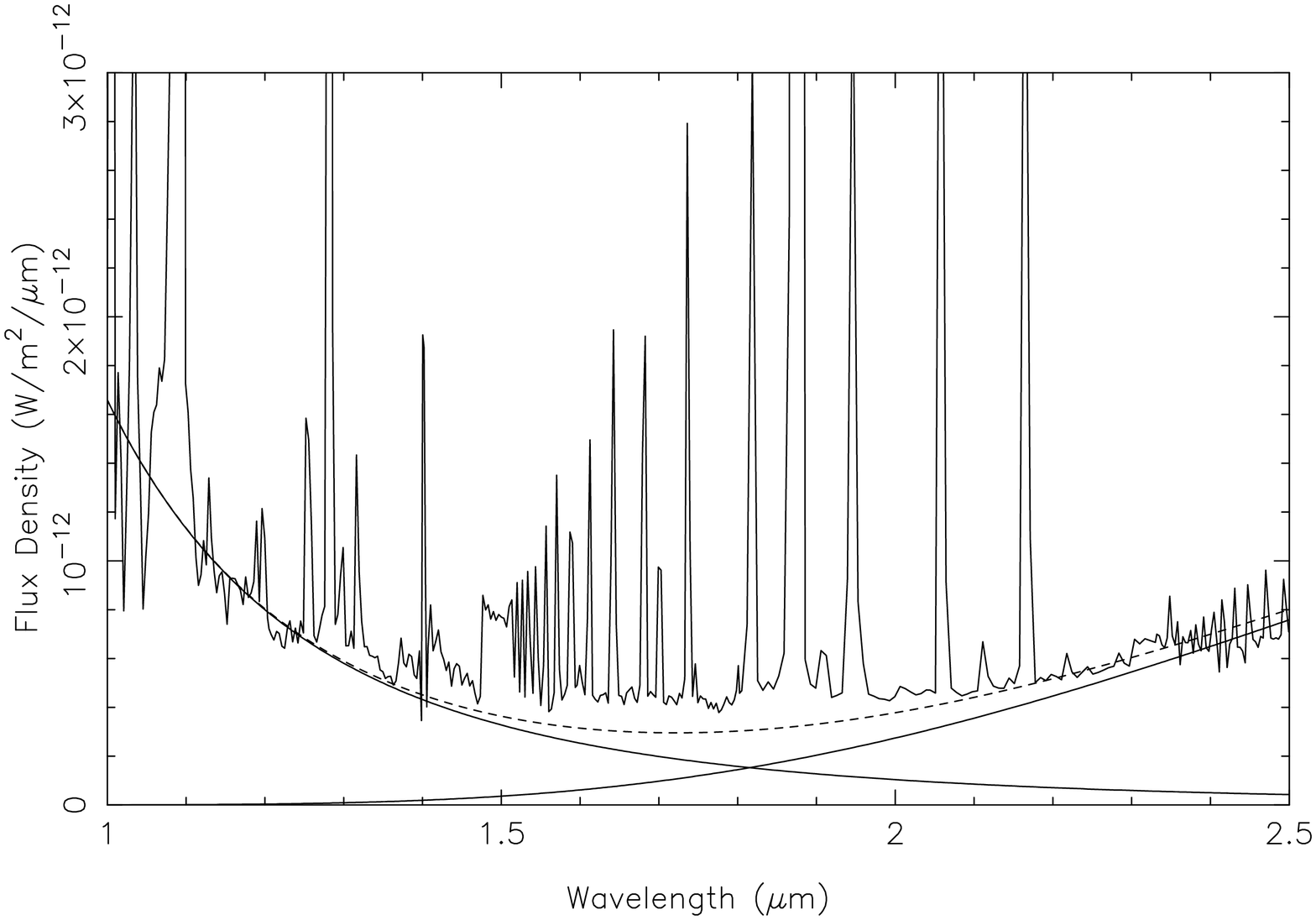}
\end{center}}\end{minipage}

\vspace*{0.3in}
\noindent{\bf Figure 4: }  
The spectrum of G45.12+0.13 after removing the free-free component shown in
figure 2, and after
correcting for extinction.  The two solid lines represent simple models
for (i) starlight (rising to shorter wavelengths) and (ii) dust
modified by a grey-body law (rising to longer wavelengths).  The dashed
line is the sum of these two components and can be seen to be a good match to
the residual continuum emission.  Note also the prominent blends of emission
lines just longwards of the Brackett and Pfund limits mentioned in
the text.

\end{document}